\documentclass[twocolumn]{aastex63}

\usepackage{amsmath}
\usepackage{amssymb}
\usepackage{bigstrut}
\usepackage{bm,color}
\usepackage{booktabs}
\usepackage{csquotes}
\usepackage{fontawesome}
\usepackage{graphicx}
\usepackage{hyperref}
\usepackage{ifthen}
\usepackage{multirow}
\usepackage{nicefrac}

\usepackage{savesym} 
\savesymbol{tablenum}
\usepackage{siunitx}
\restoresymbol{SIX}{tablenum}

\usepackage{physics}
\savesymbol{op}

\usepackage{textcomp}
\usepackage{threeparttable}
\usepackage{ulem}
\usepackage{verbatim}
\usepackage{xcolor}
\usepackage{xspace}
\MakeOuterQuote{"}

\graphicspath{{figures/}{./}} 

\newcommand{\secref}{Section\xspace}

\newcommand{\figref}{Figure\xspace}

\newcommand{\LenseFlow}{{\sc LenseFlow}\xspace}

\newcommand{\vanish}[1]{}

\newcommand{\sqdeg}{$\deg^2$\xspace}

\DeclareMathAlphabet{\mathbbgreek}{U}{bbold}{m}{n}

\newcommand{\op}[1]{\mathbb{#1}}
\newcommand{\Aphi}{A_{\phi}}
\newcommand{\Aphiarrow}{A_{\phi}^{100\rightarrow2000}}
\newcommand{\AL}{A_{\rm L}}
\newcommand{\DAL}{\Delta A_{\rm L}}
\newcommand{\Af}{A_{f}}
\newcommand{\Len}[1][]{\op L\ifthenelse{\equal{#1}{}}{}{(#1)}}
\newcommand{\Cflen}[1][]{\op{\widetilde{C}}_{f}\ifthenelse{\equal{#1}{}}{}{(#1)}}
\newcommand{\Cf}[1][]{\op C_{f}\ifthenelse{\equal{#1}{}}{}{(#1)}}
\newcommand{\Cphi}[1][]{\op C_{\phi}\ifthenelse{\equal{#1}{}}{}{(#1)}}
\newcommand{\Cn}{\op C_{n}}
\newcommand{\opG}[1][]{\op G\ifthenelse{\equal{#1}{}}{}{(#1)}}
\newcommand{\D}[1][]{\op D\ifthenelse{\equal{#1}{}}{}{(#1)}}
\newcommand{\Mpix}{\op M_{\rm p}}
\newcommand{\Mfourier}{\op M_{\rm f}}
\newcommand{\I}{\mathbbgreek{1}}

\newcommand{\mix}{\prime}
\newcommand{\Pcal}{P_{\rm cal}}

\newcommand{\psipol}{\psi_{\rm pol}}
\newcommand{\epsQ}{\epsilon_{\rm Q}}
\newcommand{\epsU}{\epsilon_{\rm U}}

\newcommand{\dsultradeep}{\textsc{100d-deep}\xspace}
\newcommand{\dsonehundred}{\textsc{100d}\xspace}
\newcommand{\dsfivehundredfull}{\textsc{500d-full}\xspace}
\newcommand{\dsfivehundredLT}{\textsc{500d-lt}\xspace}
\newcommand{\dsfivehundred}{\textsc{500d}\xspace}

\defcitealias{millea2020}{MAW20}
\defcitealias{gratton2019}{GC19}

\begin{document}
\title{Optimal CMB Lensing Reconstruction and Parameter Estimation with SPTpol Data}

\shortauthors{M.~Millea, C.~M.~Daley, T-L.~Chou, E.~Anderes, et al.}
\author[0000-0001-7317-0551]{M.~Millea} \affiliation{Department of Physics, University of California, Berkeley, CA, USA 94720}
\author{C.~M.~Daley} \affiliation{Astronomy Department, University of Illinois at Urbana-Champaign, 1002 W. Green Street, Urbana, IL 61801, USA}
\author{T-L.~Chou} \affiliation{Kavli Institute for Cosmological Physics, University of Chicago, 5640 South Ellis Avenue, Chicago, IL, USA 60637} \affiliation{Department of Physics, University of Chicago, 5640 South Ellis Avenue, Chicago, IL, USA 60637}
\author{E.~Anderes} \affiliation{Department of Statistics, University of California, One Shields Avenue, Davis, CA, USA 95616}
\author{P.~A.~R.~Ade} \affiliation{Cardiff University, Cardiff CF10 3XQ, United Kingdom}
\author{A.~J.~Anderson} \affiliation{Fermi National Accelerator Laboratory, MS209, P.O. Box 500, Batavia, IL 60510}
\author{J.~E.~Austermann} \affiliation{NIST Quantum Devices Group, 325 Broadway Mailcode 817.03, Boulder, CO, USA 80305} \affiliation{Department of Physics, University of Colorado, Boulder, CO, USA 80309}
\author{J.~S.~Avva} \affiliation{Department of Physics, University of California, Berkeley, CA, USA 94720}
\author{J.~A.~Beall} \affiliation{NIST Quantum Devices Group, 325 Broadway Mailcode 817.03, Boulder, CO, USA 80305}
\author{A.~N.~Bender} \affiliation{High Energy Physics Division, Argonne National Laboratory, 9700 S. Cass Avenue, Argonne, IL, USA 60439} \affiliation{Kavli Institute for Cosmological Physics, University of Chicago, 5640 South Ellis Avenue, Chicago, IL, USA 60637}
\author[0000-0002-5108-6823]{B.~A.~Benson} \affiliation{Fermi National Accelerator Laboratory, MS209, P.O. Box 500, Batavia, IL 60510} \affiliation{Kavli Institute for Cosmological Physics, University of Chicago, 5640 South Ellis Avenue, Chicago, IL, USA 60637} \affiliation{Department of Astronomy and Astrophysics, University of Chicago, 5640 South Ellis Avenue, Chicago, IL, USA 60637}
\author[0000-0003-4847-3483]{F.~Bianchini} \affiliation{School of Physics, University of Melbourne, Parkville, VIC 3010, Australia}
\author[0000-0001-7665-5079]{L.~E.~Bleem} \affiliation{High Energy Physics Division, Argonne National Laboratory, 9700 S. Cass Avenue, Argonne, IL, USA 60439} \affiliation{Kavli Institute for Cosmological Physics, University of Chicago, 5640 South Ellis Avenue, Chicago, IL, USA 60637}
\author{J.~E.~Carlstrom} \affiliation{Kavli Institute for Cosmological Physics, University of Chicago, 5640 South Ellis Avenue, Chicago, IL, USA 60637} \affiliation{Department of Physics, University of Chicago, 5640 South Ellis Avenue, Chicago, IL, USA 60637} \affiliation{High Energy Physics Division, Argonne National Laboratory, 9700 S. Cass Avenue, Argonne, IL, USA 60439} \affiliation{Department of Astronomy and Astrophysics, University of Chicago, 5640 South Ellis Avenue, Chicago, IL, USA 60637} \affiliation{Enrico Fermi Institute, University of Chicago, 5640 South Ellis Avenue, Chicago, IL, USA 60637}
\author{C.~L.~Chang} \affiliation{Kavli Institute for Cosmological Physics, University of Chicago, 5640 South Ellis Avenue, Chicago, IL, USA 60637} \affiliation{High Energy Physics Division, Argonne National Laboratory, 9700 S. Cass Avenue, Argonne, IL, USA 60439} \affiliation{Department of Astronomy and Astrophysics, University of Chicago, 5640 South Ellis Avenue, Chicago, IL, USA 60637}
\author{P.~Chaubal} \affiliation{School of Physics, University of Melbourne, Parkville, VIC 3010, Australia}
\author{H.~C.~Chiang} \affiliation{Department of Physics, McGill University, 3600 Rue University, Montreal, Quebec H3A 2T8, Canada} \affiliation{School of Mathematics, Statistics \& Computer Science, University of KwaZulu-Natal, Durban, South Africa}
\author{R.~Citron} \affiliation{University of Chicago, 5640 South Ellis Avenue, Chicago, IL, USA 60637}
\author{C.~Corbett~Moran} \affiliation{Jet Propulsion Laboratory, Pasadena, CA 91109, USA}
\author[0000-0001-9000-5013]{T.~M.~Crawford} \affiliation{Kavli Institute for Cosmological Physics, University of Chicago, 5640 South Ellis Avenue, Chicago, IL, USA 60637} \affiliation{Department of Astronomy and Astrophysics, University of Chicago, 5640 South Ellis Avenue, Chicago, IL, USA 60637}
\author{A.~T.~Crites} \affiliation{Kavli Institute for Cosmological Physics, University of Chicago, 5640 South Ellis Avenue, Chicago, IL, USA 60637} \affiliation{Department of Astronomy and Astrophysics, University of Chicago, 5640 South Ellis Avenue, Chicago, IL, USA 60637} \affiliation{Dunlap Institute for Astronomy \& Astrophysics, University of Toronto, 50 St George St, Toronto, ON, M5S 3H4, Canada} \affiliation{Department of Astronomy \& Astrophysics, University of Toronto, 50 St George St, Toronto, ON, M5S 3H4, Canada}
\author{T.~de~Haan} \affiliation{Department of Physics, University of California, Berkeley, CA, USA 94720} \affiliation{Physics Division, Lawrence Berkeley National Laboratory, Berkeley, CA, USA 94720}
\author{M.~A.~Dobbs} \affiliation{Department of Physics, McGill University, 3600 Rue University, Montreal, Quebec H3A 2T8, Canada} \affiliation{Canadian Institute for Advanced Research, CIFAR Program in Gravity and the Extreme Universe, Toronto, ON, M5G 1Z8, Canada}
\author{W.~Everett} \affiliation{Department of Astrophysical and Planetary Sciences, University of Colorado, Boulder, CO, USA 80309}
\author{J.~Gallicchio} \affiliation{Kavli Institute for Cosmological Physics, University of Chicago, 5640 South Ellis Avenue, Chicago, IL, USA 60637} \affiliation{Harvey Mudd College, 301 Platt Blvd., Claremont, CA 91711}
\author{E.~M.~George} \affiliation{European Southern Observatory, Karl-Schwarzschild-Str. 2, 85748 Garching bei M\"{u}nchen, Germany} \affiliation{Department of Physics, University of California, Berkeley, CA, USA 94720}
\author{N.~Goeckner-Wald} \affiliation{Dept. of Physics, Stanford University, 382 Via Pueblo Mall, Stanford, CA 94305}
\author{S.~Guns} \affiliation{Department of Physics, University of California, Berkeley, CA, USA 94720}
\author{N.~Gupta} \affiliation{School of Physics, University of Melbourne, Parkville, VIC 3010, Australia}
\author{N.~W.~Halverson} \affiliation{Department of Astrophysical and Planetary Sciences, University of Colorado, Boulder, CO, USA 80309} \affiliation{Department of Physics, University of Colorado, Boulder, CO, USA 80309}
\author{J.~W.~Henning} \affiliation{High Energy Physics Division, Argonne National Laboratory, 9700 S. Cass Avenue, Argonne, IL, USA 60439} \affiliation{Kavli Institute for Cosmological Physics, University of Chicago, 5640 South Ellis Avenue, Chicago, IL, USA 60637}
\author{G.~C.~Hilton} \affiliation{NIST Quantum Devices Group, 325 Broadway Mailcode 817.03, Boulder, CO, USA 80305}
\author[0000-0002-0463-6394]{G.~P.~Holder} \affiliation{Astronomy Department, University of Illinois at Urbana-Champaign, 1002 W. Green Street, Urbana, IL 61801, USA} \affiliation{Department of Physics, University of Illinois Urbana-Champaign, 1110 W. Green Street, Urbana, IL 61801, USA} \affiliation{Canadian Institute for Advanced Research, CIFAR Program in Gravity and the Extreme Universe, Toronto, ON, M5G 1Z8, Canada}
\author{W.~L.~Holzapfel} \affiliation{Department of Physics, University of California, Berkeley, CA, USA 94720}
\author{J.~D.~Hrubes} \affiliation{University of Chicago, 5640 South Ellis Avenue, Chicago, IL, USA 60637}
\author{N.~Huang} \affiliation{Department of Physics, University of California, Berkeley, CA, USA 94720}
\author{J.~Hubmayr} \affiliation{NIST Quantum Devices Group, 325 Broadway Mailcode 817.03, Boulder, CO, USA 80305}
\author{K.~D.~Irwin} \affiliation{SLAC National Accelerator Laboratory, 2575 Sand Hill Road, Menlo Park, CA 94025} \affiliation{Dept. of Physics, Stanford University, 382 Via Pueblo Mall, Stanford, CA 94305}
\author{L.~Knox} \affiliation{Department of Physics and Astronomy, University of California, One Shields Avenue, Davis, CA, USA 95616}
\author{A.~T.~Lee} \affiliation{Department of Physics, University of California, Berkeley, CA, USA 94720} \affiliation{Physics Division, Lawrence Berkeley National Laboratory, Berkeley, CA, USA 94720}
\author{D.~Li} \affiliation{NIST Quantum Devices Group, 325 Broadway Mailcode 817.03, Boulder, CO, USA 80305} \affiliation{SLAC National Accelerator Laboratory, 2575 Sand Hill Road, Menlo Park, CA 94025}
\author{A.~Lowitz} \affiliation{Department of Astronomy and Astrophysics, University of Chicago, 5640 South Ellis Avenue, Chicago, IL, USA 60637}
\author{J.~J.~McMahon} \affiliation{Kavli Institute for Cosmological Physics, University of Chicago, 5640 South Ellis Avenue, Chicago, IL, USA 60637} \affiliation{Department of Physics, University of Chicago, 5640 South Ellis Avenue, Chicago, IL, USA 60637} \affiliation{Department of Astronomy and Astrophysics, University of Chicago, 5640 South Ellis Avenue, Chicago, IL, USA 60637}
\author{S.~S.~Meyer} \affiliation{Kavli Institute for Cosmological Physics, University of Chicago, 5640 South Ellis Avenue, Chicago, IL, USA 60637} \affiliation{Department of Physics, University of Chicago, 5640 South Ellis Avenue, Chicago, IL, USA 60637} \affiliation{Department of Astronomy and Astrophysics, University of Chicago, 5640 South Ellis Avenue, Chicago, IL, USA 60637} \affiliation{Enrico Fermi Institute, University of Chicago, 5640 South Ellis Avenue, Chicago, IL, USA 60637}
\author{L.~M.~Mocanu} \affiliation{Kavli Institute for Cosmological Physics, University of Chicago, 5640 South Ellis Avenue, Chicago, IL, USA 60637} \affiliation{Department of Astronomy and Astrophysics, University of Chicago, 5640 South Ellis Avenue, Chicago, IL, USA 60637} \affiliation{Institute of Theoretical Astrophysics, University of Oslo, P.O.Box 1029 Blindern, N-0315 Oslo, Norway}
\author{J.~Montgomery} \affiliation{Department of Physics, McGill University, 3600 Rue University, Montreal, Quebec H3A 2T8, Canada}
\author{T.~Natoli} \affiliation{Department of Astronomy and Astrophysics, University of Chicago, 5640 South Ellis Avenue, Chicago, IL, USA 60637} \affiliation{Kavli Institute for Cosmological Physics, University of Chicago, 5640 South Ellis Avenue, Chicago, IL, USA 60637}
\author{J.~P.~Nibarger} \affiliation{NIST Quantum Devices Group, 325 Broadway Mailcode 817.03, Boulder, CO, USA 80305}
\author{G.~Noble} \affiliation{Department of Physics, McGill University, 3600 Rue University, Montreal, Quebec H3A 2T8, Canada}
\author{V.~Novosad} \affiliation{Materials Sciences Division, Argonne National Laboratory, 9700 S. Cass Avenue, Argonne, IL, USA 60439}
\author{Y.~Omori} \affiliation{Dept. of Physics, Stanford University, 382 Via Pueblo Mall, Stanford, CA 94305}
\author{S.~Padin} \affiliation{Kavli Institute for Cosmological Physics, University of Chicago, 5640 South Ellis Avenue, Chicago, IL, USA 60637} \affiliation{Department of Astronomy and Astrophysics, University of Chicago, 5640 South Ellis Avenue, Chicago, IL, USA 60637} \affiliation{California Institute of Technology, MS 249-17, 1216 E. California Blvd., Pasadena, CA, USA 91125}
\author{S.~Patil} \affiliation{School of Physics, University of Melbourne, Parkville, VIC 3010, Australia}
\author{C.~Pryke} \affiliation{School of Physics and Astronomy, University of Minnesota, 116 Church Street S.E. Minneapolis, MN, USA 55455}
\author[0000-0003-2226-9169]{C.~L.~Reichardt} \affiliation{School of Physics, University of Melbourne, Parkville, VIC 3010, Australia}
\author{J.~E.~Ruhl} \affiliation{Physics Department, Center for Education and Research in Cosmology and Astrophysics, Case Western Reserve University, Cleveland, OH, USA 44106}
\author{B.~R.~Saliwanchik} \affiliation{Physics Department, Center for Education and Research in Cosmology and Astrophysics, Case Western Reserve University, Cleveland, OH, USA 44106} \affiliation{Department of Physics, Yale University, P.O. Box 208120, New Haven, CT 06520-8120}
\author{K.~K.~Schaffer} \affiliation{Kavli Institute for Cosmological Physics, University of Chicago, 5640 South Ellis Avenue, Chicago, IL, USA 60637} \affiliation{Enrico Fermi Institute, University of Chicago, 5640 South Ellis Avenue, Chicago, IL, USA 60637} \affiliation{Liberal Arts Department, School of the Art Institute of Chicago, 112 S Michigan Ave, Chicago, IL, USA 60603}
\author{C.~Sievers} \affiliation{University of Chicago, 5640 South Ellis Avenue, Chicago, IL, USA 60637}
\author{G.~Smecher} \affiliation{Department of Physics, McGill University, 3600 Rue University, Montreal, Quebec H3A 2T8, Canada} \affiliation{Three-Speed Logic, Inc., Victoria, B.C., V8S 3Z5, Canada}
\author{A.~A.~Stark} \affiliation{Harvard-Smithsonian Center for Astrophysics, 60 Garden Street, Cambridge, MA, USA 02138}
\author{B.~Thorne} \affiliation{Department of Physics and Astronomy, University of California, One Shields Avenue, Davis, CA, USA 95616}
\author{C.~Tucker} \affiliation{Cardiff University, Cardiff CF10 3XQ, United Kingdom}
\author{T.~Veach} \affiliation{Space Science and Engineering Division, Southwest Research Institute, San Antonio, TX 78238}
\author{J.~D.~Vieira} \affiliation{Astronomy Department, University of Illinois at Urbana-Champaign, 1002 W. Green Street, Urbana, IL 61801, USA} \affiliation{Department of Physics, University of Illinois Urbana-Champaign, 1110 W. Green Street, Urbana, IL 61801, USA}
\author{G.~Wang} \affiliation{High Energy Physics Division, Argonne National Laboratory, 9700 S. Cass Avenue, Argonne, IL, USA 60439}
\author[0000-0002-3157-0407]{N.~Whitehorn} \affiliation{Department of Physics and Astronomy, Michigan State University, 567 Wilson Road, East Lansing, MI 48824}
\author[0000-0001-5411-6920]{W.~L.~K.~Wu} \affiliation{Kavli Institute for Cosmological Physics, University of Chicago, 5640 South Ellis Avenue, Chicago, IL, USA 60637} \affiliation{SLAC National Accelerator Laboratory, 2575 Sand Hill Road, Menlo Park, CA 94025}
\author{V.~Yefremenko} \affiliation{High Energy Physics Division, Argonne National Laboratory, 9700 S. Cass Avenue, Argonne, IL, USA 60439}

\correspondingauthor{M.~Millea} \email{mariusmillea@gmail.com}

\begin{abstract}  
  We perform the first simultaneous Bayesian parameter inference and optimal reconstruction of the gravitational lensing of the cosmic microwave background (CMB), using 100\,deg$^2$ of polarization observations from the SPTpol receiver on the South Pole Telescope. These data reach noise levels as low as 5.8\,$\mu$K-arcmin in polarization, which are low enough that the typically used quadratic estimator (QE) technique for analyzing CMB lensing is significantly sub-optimal. Conversely, the Bayesian procedure extracts all lensing information from the data and is optimal at any noise level. We infer the amplitude of the gravitational lensing potential to be $\Aphi\,{=}\,0.949\,{\pm}\,0.122$ using the Bayesian pipeline, consistent with our QE pipeline result, but with 17\% smaller error bars. The Bayesian analysis also provides a simple way to account for systematic uncertainties, performing a similar job as frequentist ``bias hardening,'' and reducing the systematic uncertainty on $\Aphi$ due to polarization calibration from almost half of the statistical error to effectively zero. Finally, we jointly constrain $\Aphi$ along with $A_{\rm L}$, the amplitude of lensing-like effects on the CMB power spectra, demonstrating that the Bayesian method can be used to easily infer parameters both from an optimal lensing reconstruction and from the delensed CMB, while exactly accounting for the correlation between the two. These results demonstrate the feasibility of the Bayesian approach on real data, and pave the way for future analysis of deep CMB polarization measurements with SPT-3G, Simons Observatory, and CMB-S4, where improvements relative to the QE can reach 1.5 times tighter constraints on $\Aphi$ and 7 times lower effective lensing reconstruction noise.
\end{abstract}

\keywords{cosmic background radiation - cosmological parameters - gravitational lensing}


\section{Introduction}

Gravitational lensing of the cosmic microwave background (CMB) occurs as CMB photons traveling to us from the last scattering surface are deflected by the gravitational potentials of intervening matter. This effect has been detected with high significance, allowing inference of the line-of-sight projected gravitational field of the intervening matter and of the late-time expansion history and geometry of the universe \citep{lewis2006a,planck2018lensing}. Better measurements of the lensing effect are one of the main goals of nearly all future CMB probes, and can help constrain dark matter, neutrinos, modified gravity, and a wealth of other cosmological physics \citep{benson2014,abazajian2016,thesimonsobservatorycollaboration2019}. 

Traditionally, analysis of lensed CMB data has relied on the so-called quadratic estimate (QE) of the gravitational lensing potential, $\phi$ \citep{zaldarriaga1999,hu02}. The QE is a frequentist point estimate of $\phi$ formed from quadratic combinations of the data. It is conceptually simple and near minimum-variance at noise levels up to and including many present day experiments. However, it was realized by \cite{hirata2003,hirata2003a} and \cite{seljak2004} that when instrumental noise levels drop below ${\sim}\,5\,\mu \rm K\,arcmin$, where lensing-induced $B$-modes begin to be resolved with signal-to-noise greater than one, the QE ceases to be minimum-variance and better analysis can extract more information from the same data. \cite{hirata2003a} were the first to construct a better estimator, using a method based on the Bayesian posterior for CMB lensing. This included a maximum a posteriori (MAP) estimate of $\phi$ which has lower variance than the QE\footnote{The MAP $\phi$ estimate from \cite{hirata2003a} has sometimes been called the ``iterative quadratic estimate,'' but because several methods exist which involve iterating something akin to a quadratic estimate, we do not use this term and instead more precisely refer to individual methods.}, and a maximum likelihood estimate (MLE) of the power spectrum of gravitational lensing potential, $C_\ell^{\phi\phi}$. These results used a number of simplifying approximations, including perfectly white noise and periodic flat-sky boundaries with no masking in the pixel domain. Extending this original work, \cite{carron2017} upgraded this MAP $\phi$ procedure to work without these approximations, rendering it applicable to realistic instrumental conditions.

Although estimates of the $\phi$ maps are useful, here we are interested in reconstructing not only $\phi$ but its theory power spectrum as well. A common misconception is that once one has a better estimate of $\phi$ (e.g. a MAP $\phi$ estimate), one can take its power spectrum, subtract a noise bias, and obtain the desired estimate of $C_\ell^{\phi\phi}$. While this does work for the QE, it is only because the QE can be analytically normalized and its power spectrum analytically noise debiased (up to some usually minor Monte Carlo corrections), yielding an unbiased estimate of the theory lensing spectrum. However, this is not generically the case for MAP estimates, for which analytic calculations of normalization and noise biases do not exist. In theory, one could try computing these entirely via Monte Carlo, but this can only be done at a single fiducial cosmological model, and it is unknown to what extent these could be cosmology-dependent or how one might deal with this. If a frequentist estimate is nevertheless desired, a more promising approach may be something akin to the $C_\ell^{\phi\phi}$ MLE proposed by \cite{hirata2003a}. However, this has not yet been demonstrated on realistic data.

An alternate approach is based on direct Bayesian inference of cosmological quantities of interest, without the need for explicit normalization and debiasing of any intermediate power spectra. Recent progress was presented in \cite{anderes2015}, who developed a Monte Carlo sampler of the Bayesian posterior of unlensed CMB temperature maps and $\phi$ maps given fixed cosmological parameters. \cite{millea2019} began the process of incorporating polarization into this procedure, resulting in a joint MAP estimate of both the $\phi$ map and the CMB polarization fields. Finally, \cite{millea2020} (hereafter \citetalias{millea2020}) extended this to a full Monte Carlo sampler and included cosmological parameters in the sampling, giving the key ingredients needed for the work here. By virtue of directly mapping out the Bayesian posterior for these quantities, this method achieves the goal of fully extracting cosmological information from lensed CMB data and is optimal at all noise levels.

Instrumental noise levels which are low enough at the relevant scales to necessitate anything beyond the QE have only recently been attained. The POLARBEAR collaboration performed the first (and to-date only) beyond-QE analysis of real data \citep{adachi2019}. This used the \cite{carron2017} MAP $\phi$ estimate to internally ``delense'' the data, removing the lensing-induced $B$-mode polarization. Unlike generic $C_\ell^{\phi\phi}$ estimation, $B$-mode delensing does not require renormalizing the $\phi$ estimate, and noise biases can be mitigated via the ``overlapping $B$-mode deprojection'' technique. 

In this work, we go a step further and perform an optimal lensing reconstruction and full parameter extraction from the lensing potential and from internally delensed bandpowers. Although similar in spirit, our methodology is quite different, however, and based on the \citetalias{millea2020} Bayesian sampling procedure rather than on any point estimates. We use the deepest 100\,deg$^2$ of South Pole Telescope polarization data obtained with the SPTpol receiver, restricting ourselves to just this deepest patch since we are mainly interested in the low-noise regime where the Bayesian procedure will outperform the QE. We infer cosmological parameters $\Aphi$ and $\AL$, along with a host of systematics parameters. The $\Aphi$ parameter is a standard parameter scaling the theory lensing spectrum as $C_\ell^{\phi\phi}\,{\rightarrow}\,\Aphi C_\ell^{\phi\phi}$. $\Aphi$ can be considered a proxy for any physical parameter that is constrained by the lensing potential, such as the matter density or the sum of neutrino masses. We choose to estimate $\Aphi$ here for simplicity, but in the future the method could easily be extended to estimate more physical parameters instead. The $\AL$ parameter scales the lensing-like contribution to the model CMB power spectrum, and is defined such that $\AL\,{=}\,1$ if the underlying cosmological model is correct. Unlike frequentist estimates, the Bayesian procedure requires a self-consistent data model which includes both $\Aphi$ and $\AL$, and we develop one here. Finally, we include several systematics parameters, noting that it is particularly easy to incorporate systematic errors in the Bayesian approach. The final output of this procedure is a Monte Carlo Markov Chain (MCMC) composed of samples of these parameters along with samples of the $\phi$ maps and unlensed CMB polarization maps, for a total of 202,808 dimensions sampled. Ultimately, we demonstrate a 17\% improvement of the Bayesian constraint on $\Aphi$ as compared to the QE. 

The results here are new in three regards:
\begin{itemize}
  \item First time a parameter ($\Aphi$) is estimated from an optimal lensing reconstruction. 
  \item First joint inference of parameters controlling the lensing potential ($\Aphi$) and controlling the CMB bandpowers ($\AL$), while fully and exactly accounting for correlation between the reconstruction and the delensed CMB.
  \item First application of a fully Bayesian method to CMB lensing data.
\end{itemize}
These demonstrate important pieces of the type of fully optimal beyond-QE analysis which will be a requirement if next-generation experiments such as SPT-3G, Simons Observatory, and CMB-S4 are to reach their full (and expected) potential \citep{benson2014,abazajian2016,thesimonsobservatorycollaboration2019}. 

The organization of the paper is as follows. The reader who wishes to skip the details of the MCMC sampling procedure and simply trust that it yields samples from the exact CMB lensing posterior can jump to the main results in Sec.~\ref{sec:results} and discussion in Sec.~\ref{sec:conclusion}. The earlier sections give the technical details of the data modeling and sampling. In Sec.~\ref{sec:data}, we describe the data and simulations used in this work. These data have been previously vetted in \cite{story15} and \cite{wu2019}, and we refer the reader to these works for various null tests, here choosing instead to concentrate on the lensing analysis. Most of the focus of this work is on the Bayesian pipeline in particular, and Sec.~\ref{sec:modeling} lays out the forward model necessary to construct the posterior for CMB lensing given the SPT data. Sec.~\ref{sec:lensing} describes the Bayesian and QE lensing pipelines, and Sec.~\ref{sec:validation} provides validation of the procedures, including on a suite of realistic simulations of the actual data.

\section{Data and simulations}
\label{sec:data}

\subsection{Data}

In this work, we use data from the 150 GHz detectors from the SPTpol receiver on the South Pole Telescope \citep{padin08,carlstrom11,bleem2012}. SPTpol has employed three different scan strategies for the observations which comprise our final dataset.

From March 2012 to April 2013, SPTpol observed a 100 deg$^2$ patch of sky ($10^\circ \times 10^\circ$) centered at right ascension (R.A.) 23$^h$30$^m$ and declination (dec.) $-55^\circ$. All observations of this field were made using an azimuthal ``lead-trail'' scan strategy, where the 100\,\sqdeg field is split into two equal halves in R.A., a ``lead'' half-field and a ``trail'' half-field. The lead half-field is observed first, followed immediately by a trail half-field observation, such that the lead and trail observations occur in the same azimuth-elevation range. Each half-field is observed by scanning the telescope in azimuth right and left across the field and then stepping up in elevation. This lead-trail strategy enables removal of ground pickup. We will refer to these data as the \dsonehundred observations.

From April 2013 to May 2014, SPTpol observed a 500\,\sqdeg patch of sky, extending from 22$^h$ to 2$^h$ in R.A. and from $-65^\circ$ to $-50^\circ$ in dec. Observations during this time were also made using the ``lead-trail'' scan strategy, and we will refer to them as the \dsfivehundredLT observations. 

From May 2014 to Sep 2016, while observing the same 500\,\sqdeg field, SPTpol switched to the ``full-field'' scan strategy in order to increase sensitivity to larger scales on the sky. In this case, constant-elevation scans are made across the entire range of R.A. of the field. We will refer to these data as the \dsfivehundredfull observations.

Our final dataset comprises 6262 \dsonehundred observations, 858 \dsfivehundredLT observations, and 3370 \dsfivehundredfull observations. Each observation records the time-ordered data (TOD) of each detector, and these TOD are filtered and calibrated before being binned into maps. Our data reduction largely follows previous TE/EE power spectrum analyses, namely \cite{crites15} for the \dsonehundred observations, and \cite{henning18} for the \dsfivehundredLT and \dsfivehundredfull observations. Here we only highlight relevant aspects for this analysis.

For the \dsonehundred observations, we use slightly different TOD filters compared to previous analysis of these data in \cite{crites15}. We subtract a 5th-order Legendre polynomial from the TOD of each detector, and then apply a high-pass filter at 0.05\,Hz, in order to match the filter choices for \dsfivehundred observations. Based on the size of our map pixels, we apply a low-pass filter at a TOD frequency corresponding to an effective $\ell\,{=}\,5000$ for anti-aliasing along the scan direction. Electrical cross-talk between detectors could bias our measurement, and in \cite{crites15} we applied the cross-talk correction to the power spectra at the end of the analysis. However, in this analysis we correct cross-talk at the TOD level by measuring a detector-to-detector cross-talk matrix, in the same way as described in \cite{henning18}.

For the \dsfivehundredLT observations, we slightly modify the filters as compared to \cite{henning18} as well. We subtract a 3rd-order Legendre polynomial from each detector's TOD, and then apply a high-pass filter at $\ell\,{=}\,100$ to further suppress atmospheric noise. We also apply a low-pass filter at $\ell\,{=}\,5000$ for anti-aliasing. For the \dsfivehundredfull observations, while using the same high-pass and low-pass filters, we subtract a 5th-order Legendre polynomial instead, due to each scan being twice as long in the scan direction. Electrical cross-talk is corrected as described in \cite{henning18}. 

The TOD of each detector are calibrated relative to one another using an internal thermal source and observations of the Galactic HII region RCW38. The polarization angles of each detector are calibrated by observing an external polarized thermal source, as described in \cite{crites15}. We bin detector TOD into maps with square $1^\prime$ pixels using the oblique Lambert azimuthal equal-area projection, centered at the \dsonehundred field center. Because the Bayesian analysis is computationally intensive and scales with the number of pixels, it is advantageous to reduce the number of pixels in the final data map as much as possible. Since our analysis does not use modes above $\ell_{\rm max}\,{=}\,3000$, we can losslessly downgrade the data maps to $3^\prime$ arcmin pixels, for which the Nyquist frequency is $\ell_{\rm nyq}\,{=}\,3400$. Downgrading is performed by first applying an anti-aliasing isotropic low-pass at $\ell_{\rm nyq}$, averaging pixels together, then deconvolving the pixel-window function to match the original $1^\prime$ map (the remaining $1^\prime$ pixel-window function is accounted for in our forward model for the data). The reason for not making maps directly at $3^\prime$ resolution is because the anti-aliasing filter is most easily applied to the intermediate $1^\prime$ maps, rather than at the TOD level.

Because we are interested in a low-noise data set where the improvement over the QE is most evident, we only run the analysis on data within the \dsonehundred footprint, and only on polarization data. The final data product is a set of coadded $260{\times}260$ pixel $Q$ and $U$ maps. The effective noise level of the \dsultradeep dataset inside the mask used in the analysis is 6.0\,$\mu$K-arcmin in polarization over the multipole range $1000\,{<}\,{\ell}\,{<}\,3000$, dipping to 5.8\,$\mu$K-arcmin in the deepest parts of the field.

\section{Modeling}
\label{sec:modeling}

\begin{figure*}
  \includegraphics[width=\textwidth]{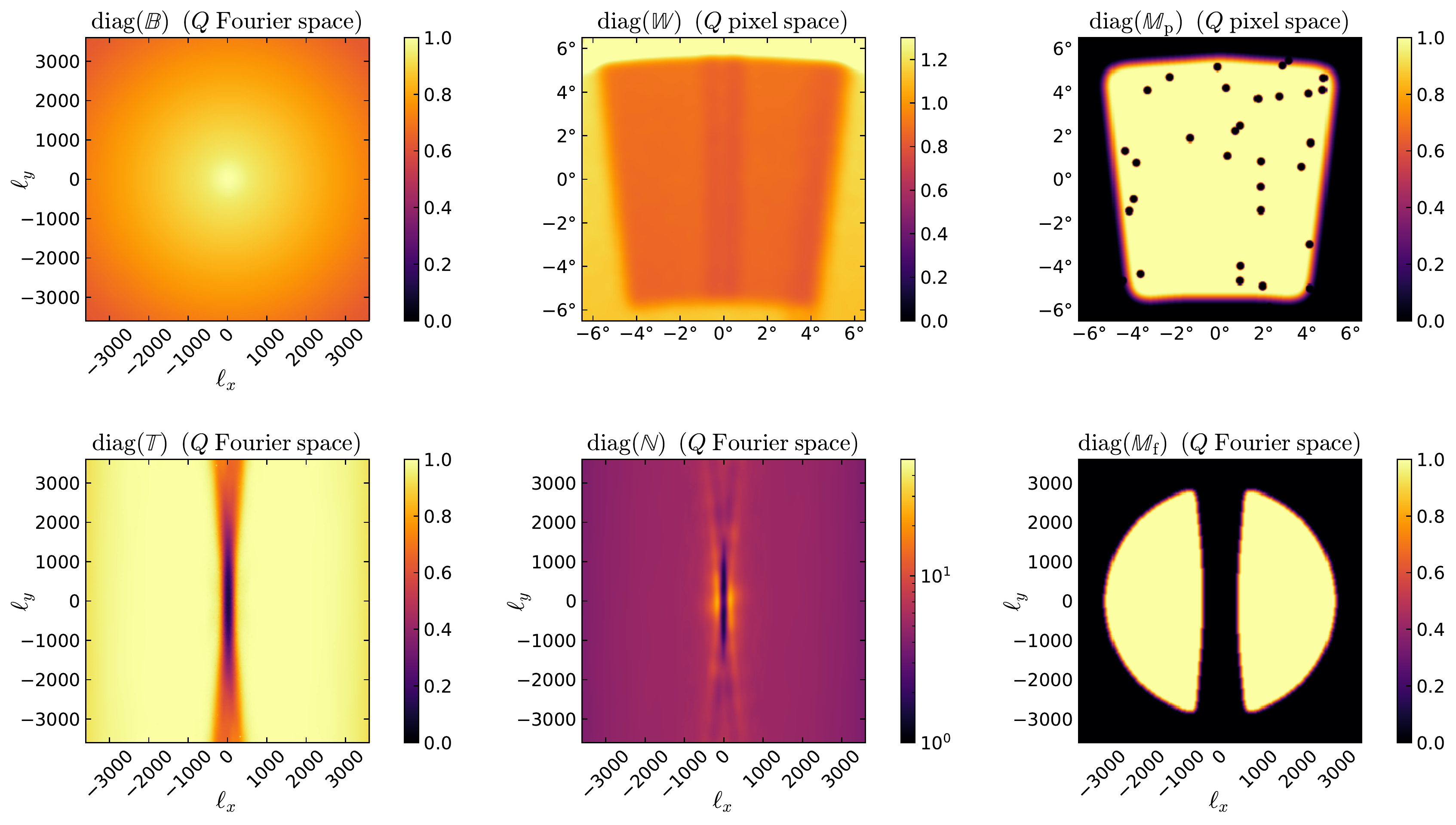}
  \caption{To help orient the reader, a visualization of the various linear operators which enter the CMB lensing posterior in Eq.~\eqref{eq:jointposterior}. The operators $\op{B}$ and $\op{T}$ are the beams and transfer functions, $\op{W}$ and $\op{N}$ together form the noise covariance as $\Cn\,{=}\,\op{W}\,\op{N}\,\op{W}^\dagger$, and $\Mpix$ and $\Mfourier$ are the pixel-space and Fourier-space masks, respectively (see Sec.~\ref{sec:modeling} for a full description). These operators correspond to $N_{\rm pix}\,{\times}\,N_{\rm pix}$ matrices which act on the $N_{\rm pix}$-dimensional vector space of spin-2 (i.e. polarization) 2D maps or 2D Fourier transforms (here $N_{\rm pix}\,{=}\,2\,{\cdot}\,260^2$). The quantities plotted above are the $Q$ component of the diagonal of these matrices when represented in the basis labeled in each plot. For $\op{B}$, $\Mpix$, and $\Mfourier$, the $Q$ and $U$ components are taken to be identical, while for $\op{T}$, $\op{W}$, and $\op{N}$, they are allowed to be different (but qualitatively end up very similar, and hence only $Q$ is shown).}
  \label{fig:operators}
\end{figure*}

To compute the Bayesian posterior for CMB lensing, we require a forward data model and a set of priors. The data, $d$, which is used as input to the Bayesian pipeline, is a masked and ``noise-filled'' version of the $QU$ data produced by the map-making described in the previous section (we will describe the masking and what we mean by noise-filled later in this section). The model we assume for $d$ and later demonstrate is sufficiently accurate is
\begin{multline}
  \label{eq:datamodel}
  d = \op \Mfourier \, \Mpix \, \op{R}_{\rm obs} \times  \\
  \, \big[ \Pcal \, \op{R}(\psipol) \, \op{T} \, \op{B}(\beta_i) \, \Len[\phi] \, f + \epsQ t_{\rm Q} + \epsU t_{\rm U} \big] + n
\end{multline}
where
\begin{itemize}
  \item $f$ are the unlensed CMB polarization fields,
  \item $\phi$ is the gravitational lensing potential,
  \item $n$ is the instrumental and/or atmospheric noise,
  \item $\Len[\phi]$ is the lensing operation,
  \item $\op B(\beta_i)$ is the beam smoothing operation, controlled by a set of beam eigenmode amplitudes, $\beta_i$,
  \item $\op T$ are the transfer functions,
  \item $\op{R}(\psi_{\rm pol})$ is a global Q/U rotation by an angle $\psi_{\rm pol}$, representing the absolute instrumental calibration,
  \item $\op{R}_{\rm obs}$ is a fixed but spatially dependent Q/U rotation which aligns the flat-sky Q/U basis vectors to the data observation basis, the inverse of the operation sometimes referred to as ``polarization flattening'',
  \item $\Pcal$ is the polarization calibration parameter,
  \item $t_{\rm Q/U}$ are temperature-to-polarization monopole leakage templates and $\epsilon_{\rm Q/U}$ are their amplitude coefficients,
  \item $\Mpix$ and $\Mfourier$ are pixel-space and Fourier-space masking operations, respectively.
\end{itemize}
We use the notation that lower-case regular letters represent maps, and double-struck upper-case letters represent linear operators on the $N_{\rm pix}$-dimensional abstract vector space spanned by all possible maps. Later in the paper, we also use the notation that Diagonal($x$) refers to a diagonal matrix with the vector $x$ along the diagonal, and diag($\op{A}$) returns the vector along the diagonal of the matrix $\op{A}$. 

We adopt Gaussian priors on the fields $f$, $\phi$, and $n$
\begin{align}
\label{eqn:fprior}
f &\sim \mathcal N\big(0, \Cf[\Af]\big) \\ 
\label{eqn:phiprior}
\phi &\sim \mathcal N\big(0, \Cphi[\Aphi]\big) \\
\label{eqn:nprior}
n &\sim \mathcal N\big(0, \Cn\big),
\end{align}
where $\Cf[\Af]$, $\Cphi[\Aphi]$, and $\Cn$ denote the covariance operators for unlensed CMB polarization, the lensing potential, and the experimental noise. The first two depend on parameters which control the amplitude of the overall power spectra, 
\begin{align}
\label{eq:Cfr}
\Cf[\Af] &= \Af \Cf^0 \\
\Cphi[\Aphi] &= \Cphi^0 + (\Aphi-1)\,\op{V}\,\Cphi^0.
\end{align}
where $\Cf^0$ and $\Cphi^0$ are evaluated at the best-fit {\it Planck} cosmology. The lensing amplitude parameter, $\Aphi$, is the main cosmological parameter of interest in this work, and scales the amplitude of the fiducial lensing potential within some window, $\op{V}$. The window allows us to estimate the amplitude just within a given multipole range, which here we take to be $\ell=(100,2000)$ to match previous SPT lensing analyses. This parameter is sometimes denoted as $\Aphiarrow$, but throughout this work, unless otherwise stated or included for clarity, we will drop the superscript and simply refer to
\begin{align}
  \Aphi\,{\equiv}\,\Aphiarrow.
\end{align}
The unlensed CMB amplitude parameter, $\Af$, functions as a proxy for the {\it Planck} absolute calibration, and allows us to marginalize over the uncertainty in this quantity. Incorporating the $\AL$ parameter is slightly less straightforward than either $\Aphi$ or $\Af$, and this discussion is delayed until Sec.~\ref{sec:AphiAL}. All other cosmological parameters not explicitly sampled are assumed to be perfectly known and fixed their true value given the fiducial model. 

We assume uniform priors on the cosmological and instrumental parameters: $\Af$, $\Aphi$, $\Pcal$, $\psipol$, $\epsQ$, and $\epsU$, and unit normal priors on the $\beta_i$ (discussed in Sec.~\ref{sec:beams}). 

This set of choices fully specifies the posterior distribution over all variables, given in Eq.~\eqref{eq:jointposterior}:
\begin{widetext}
\begin{multline}
\label{eq:jointposterior}
\mathcal{P}\big(\,f,\phi,\Aphi,\Af,\Pcal,\psipol,\epsQ,\epsU,\beta_i\,|\,d\,\big) \;\propto \\[8pt]
\frac{\exp\left\{ -\cfrac{\left[d - \op \Mfourier \, \Mpix \, \op{R}_{\rm obs} \, \, \big( \Pcal \, \op{R}(\psipol) \, \op{T} \, \op{B}(\beta_i) \, \Len[\phi] \, f + \epsQ t_{\rm Q} + \epsU t_{\rm U} \big) \right]^2}{2 \,\Cn} \right\}}{\det \Cn^{\nicefrac{1}{2}}} \;
\frac{\exp\left\{ -\cfrac{f^2}{2\,\Cf[\Af]} \right\}}{\det  \Cf[\Af]^{\nicefrac{1}{2}}} \;
\frac{\exp\left\{ -\cfrac{\phi^2}{2\,\Cphi[\Aphi]} \right\}}{\det  \Cphi[\Aphi]^{\nicefrac{1}{2}}} \;
\mathcal{P}(\beta_i) \\[15pt]
\hfill \textrm{where we use the shorthand} \; x^2/\op N \equiv x^\dagger \op N^{-1} x \; \textrm{here and throughout the paper}.
\end{multline}
\end{widetext}
Following the terminology of \citetalias{millea2020}, we refer to this as the ``joint posterior,'' in contrast to the ``marginal posterior'' which would analytically marginalize out $f$. 

\subsection{Calibration}
\label{sec:Pcal}

Performing a change-of-variables from $f \rightarrow f/\sqrt\Af$ in Eq.~\eqref{eq:jointposterior} makes it clear that the posterior constrains only the product $\Pcal\sqrt\Af$. Thus, without loss of generality, we fix $\Af\,{=}\,1$ in our sampling and only explicitly sample the $\Pcal$ parameter. The resulting constraints on $\Pcal$ can be interpreted as a constraint on $\Pcal\sqrt\Af$, or equivalently as a constraint on the SPT polarization calibration when calibrating to a perfectly known theory unlensed CMB spectrum given by the {\it Planck} best-fit. 

An estimate of $\Pcal$ can be obtained by comparing SPTpol E maps with those made by \textit{Planck}. For the \dsfivehundred data, \cite{henning18} measured $\Pcal\,{=}\,1.06$, and for the \dsonehundred data, \cite{crites2015} measured $\Pcal\,{=}\,1.048$. A weighted combination of the two predicts $\Pcal\,{\sim}\,1.055$ for the \dsultradeep data.

This external estimate of $\Pcal$, however, is not directly used, because we do not correct the raw data by a best-fit $\Pcal$. Instead, we include $\Pcal$ in the forward model for the data and sample its value in our MCMC chains. Note that this approach is unique for a lensing analysis, because it means that the calibration is jointly estimated at the same time as other systematics, at the same time as cosmological parameters, and even at the same time as the reconstructed $\phi$ maps themselves. We will see in Sec.~\ref{sec:systematics} that this has concrete benefits, mainly that it reduces the impact of the uncertainty on $\Pcal$ on the final cosmological uncertainty. As a consistency check, we will also show that the range of $\Pcal$ values allowed by the MCMC chain is consistent with $\Pcal\,{\sim}\,1.055$. 

For the QE pipeline where there is no analogous approach, we do correct the data, however we correct by the best-fit value from the Bayesian pipeline for easier comparison between the two. All of the systematics parameters described in the following sub-sections are handled in the same way as $\Pcal$, by sampling in the Bayesian case and by applying a best-fit correction in the QE case.

\subsection{Global polarization angle}\
\label{sec:psipol}

Assuming negligible foregrounds and a non parity-violating cosmological model, we expect the cross-spectra between $TB$ and $EB$ to be consistent with zero. A systematic error in the global polarization angle calibration of the instrument, $\psipol$, can also create a signal in these channels. A typical approach is to determine $\psipol$ by finding the value that nulls the $TB$ and $EB$ channels \citep{keating2012}. This was the approach taken in \cite{wu2019} for a subset of the same data used here, which found $\psipol\,{=}\,0.63^\circ\,{\pm}\,0.04^\circ$. 

We include the global polarization rotation in the forward data model in the form of the operator $\op{R}(\psipol)$, and jointly infer $\psipol$ along with the other systematics and cosmological parameters. Because the prior on $f$ assumes no correlation between $EB$ (i.e. $\Cf$ is diagonal in $EB$ Fourier space), the MCMC chain will implicitly try to find the $\psipol$ which nulls the $EB$ channel. As we will see in Sec.~\ref{sec:consistency}, the value we find is consistent with the determination from \cite{wu2019}.

\subsection{Temperature-to-polarization leakage}
\label{sec:leak}

Because the measured polarization signal effectively comes from differencing the measured intensity along two different polarization axes, any systematic mismatch affecting just one of the axes can leak the CMB temperature signal into polarization. Depending on the nature of the mismatch, different functions of the temperature map can be leaked into $Q$ and $U$. For example, a gain variation between detectors will leak a copy of the $T$ map directly, whereas pointing errors, errors in the beam width, or beam ellipticity will leak higher-order gradients of the $T$ map \citep{ade2015}. Because the temperature map is measured with very high signal-to-noise, the presence of leakage can be detected by cross correlating temperature and $Q$ or $U$ maps (this correlation should be zero on average for the true CMB, given a Fourier mask with appropriate symmetries). Additionally, if any correlation is detected, it can simply be subtracted given an appropriate amplitude.

For the \dsultradeep data, cross correlating with the appropriate templates demonstrates that only gain-type leakage exists at appreciable levels in the maps. This type leads to a leakage of the form, 
\begin{align}
  \begin{pmatrix}
    Q \\ U 
    \end{pmatrix}	\rightarrow
  \begin{pmatrix}
    Q \\ U 
  \end{pmatrix}	+
  \begin{pmatrix}
    \epsQ T \\ \epsU T
  \end{pmatrix} 
\end{align}
where $\epsQ$ and $\epsU$ are coefficients which capture the total leakage to each channel. Minimizing the $TQ$ and $TU$ cross-correlation yields best-fit values of
\begin{align}
  \epsQ = 0.010 \;\;\;\;\;\; \epsU = 0.006.
  \label{eq:epsQU}
\end{align}
As for the other systematics, these values are only used as a consistency check, and instead the leakage templates are included in the forward model and $\epsQ$ and $\epsU$ are sampled. For convenience, we also define the spin-2 polarization fields, $t_{\rm Q}\,{\equiv}\,(T,0)$ and $t_{\rm U}\,{\equiv}\,(0,T)$, which allow writing the leakage contribution in the form seen in Eq.~\eqref{eq:jointposterior}. Finally, we note that the coefficients are small enough that no $T$ noise is introduced in the deprojection or marginalization over the leakage templates, thus the $T$ field can be taken as a fixed truth given by the measurement and does not need to be additionally sampled. As we will see in Sec.~\ref{sec:consistency}, the values preferred by the chain are in agreement with Eq.~\eqref{eq:epsQU}.

\subsection{Beams}
\label{sec:beams}

For the \dsonehundred field, the beam window function and error covariance are measured using eight independent observations of Mars. The beam in the field observations are further broadened by pointing jitter, which we estimate by making a second beam measurement using bright point sources in the \dsonehundred field, and convolving it with the Mars-derived beam. Full details can be found in \cite{crites15}. For the \dsfivehundred field, the beam is measured using seven independent Venus observations, and pointing jitter is convolved in the same way as above. Full details can be found in \cite{henning18}, where a cross-check is also performed by comparing with \textit{Planck} beams and maps. The \dsultradeep beam is computed by averaging over beam-convolved simulations of the \dsonehundred and \dsfivehundred fields, combined given the appropriate weights.

The forward data model includes the beam uncertainty in the form of a beam operator parameterized by free beam eigenmode amplitudes:
\begin{align}
  \op{B}(\beta_i) = \op{B}_0 + \beta_1 \op{B}_1 + \beta_2 \op{B}_2 + ...
\end{align}
where $\op{B}_0$ is the best-fit beam, the $\beta_i$ are beam eigenmode amplitudes, and the $\op{B}_i$ are the perturbations to the beam operator determined from an eigenmode decomposition of the beam covariance matrix. An image of $\op{B}_0$ is shown in the top left panel of Fig.~\ref{fig:operators}. We normalize the $\op{B}_i$ such that the $\beta_i$ have unit normal priors, which are included in the sampling. We keep three eigenmodes in the chain. As we will see in Sec.~\ref{sec:systematics}, none are appreciably constrained beyond their prior, indicating that the data is consistent with the fiducial beam determination. 

\subsection{Masking}
\label{sec:masking}

Our analysis applies a pixel mask, $\Mpix$, which selects the \dsultradeep field and masks bright discrete sources. The mask border is built by thresholding the noise pixel variance at 5 times its minimum value, straightening the resulting edge with a smoothing filter, and finally applying a 1\,deg$^2$ cosine apodization window. The source mask is composed of known galaxy clusters \citep{vanderlinde2010}, and point sources detected in temperature with fluxes greater than 50\,mJy \citep{everett2020}. In total, the effective sky fraction left unmasked is $99.9\,\deg^2$. This  pixel mask is shown in the top-right panel of Fig.~\ref{fig:operators}. 

We note that neither Bayesian nor QE pipelines require that the mask be apodized. However, while the Bayesian pipeline remains optimal for any mask, hard mask edges can lead to larger Monte Carlo corrections and slight sub-optimalities in the QE pipeline. To facilitate a fairer comparison, we have chosen to use apodization in the baseline case, but also present results with an unapodized mask in Sec.~\ref{sec:consistency}.

In the Fourier domain, we apply a Fourier-space mask, $\Mfourier$, shown in the bottom-right panel of Fig.~\ref{fig:operators}. The center part of the mask is built by thresholding the 2D transfer function at 0.9 to remove modes, mainly in the $\ell_x$ direction, which are significantly affected by the TOD filtering and for which the approximation that $\op{T}$ is diagonal in $QU$ Fourier space breaks down. We additionally apply an $\ell_{\rm max}\,{=}\,3000$ upper bound to limit the possible contamination from polarized extra-galactic point sources. Although there is not much information beyond $\ell\,{=}\,3000$ at these noise levels, we note that this choice is likely quite conservative and can probably be significantly relaxed in the future.

The total masking operator is chosen as $\op{M}\,{=}\,\Mfourier\,\Mpix$, i.e. pixel masking happens first. To produce the data which is input to the Bayesian pipeline, $d$, we apply $\op{M}$ to the raw data map that is output by the map making procedure. We then also self-consistently include $\op{M}$ in the data model itself. Because $\Mfourier$ and $\Mpix$ do not commute exactly, there is some small leakage of masked Fourier modes into $d$. Our analysis features a fairly conservative $\Mfourier$ and it is not a problem that the effective Fourier mask leaks slightly into the region which is formally masked by $\Mfourier$, specifically by around $\Delta\ell\,{\sim}\,10$ (set by the width of the mask kernel window function). For future analyses where a more precise cut might be desired, one could fully remove any leakage by directly deprojecting the undesired modes from the data and including the deprojection operator in the data model.

\subsection{Transfer functions}
\label{sec:tf}

\begin{figure}
  \includegraphics[width=\columnwidth]{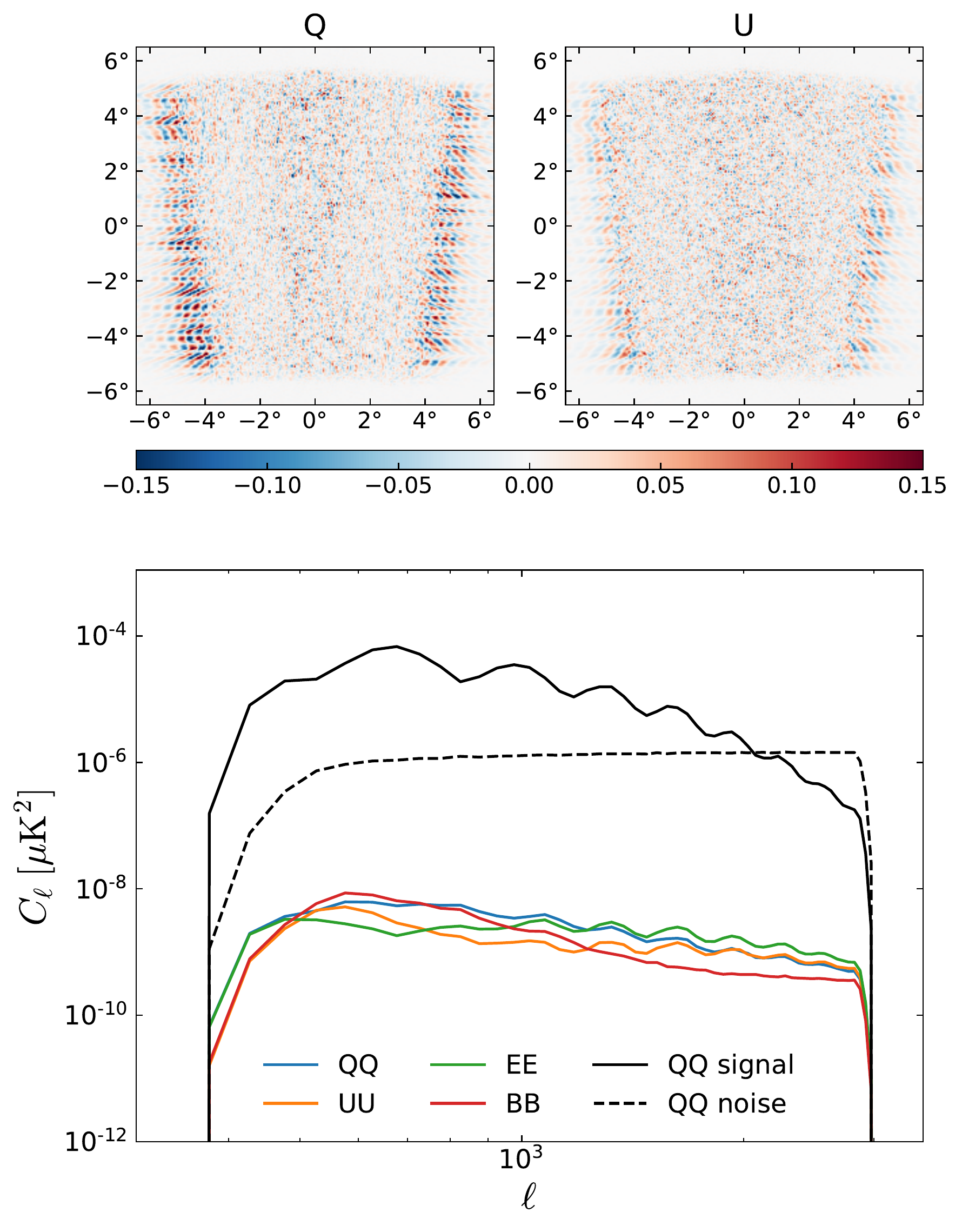}
  \caption{Validation of the approximations underlying our estimate of the transfer function, $\op{T}$ (see Sec.~\ref{sec:tf}). The top plots shows the $Q$ and $U$ components of the difference between 1) a full \dsultradeep TOD-level noise-free pipeline simulation and 2) a simple projection of the same realization then multiplication by $\op{T}$. The differences arise from mode coupling induced by the TOD filtering and Monte Carlo error in the transfer function estimation procedure. The bottom plot shows the power spectrum of these difference maps, averaged over several realizations, as well as of the $QQ$ signal and noise for comparison. Differences are 1-4 orders of magnitude below the noise power spectrum, hence negligible. We note that in both top and bottom plots, the full Fourier and pixel mask, $\op{M}$, has been applied, so as to pick out the modes which are actually relevant in the analysis.}
  \label{fig:validate_TF}
\end{figure}

The filters applied to the TOD during map making imprint an effective transfer function on the data maps, dependent on the scanning strategy and filtering choices made for each type of observation. We approximate these transfer functions, $\op{T}$, as diagonal in $QU$ Fourier space, and estimate them, as well as validating the approximation, with a set of full pipeline simulations. The full pipeline simulations are fairly computationally costly, and we take two steps to reduce the cost of this step of the analysis: 1) we simplify each simulation by reducing the number of individual observations which are included, and 2) we reduce the total number of simulations needed from ${\sim}\,400$ to only $20$ using a variance canceling technique. 

The full pipeline simulations start with a Gaussian realization of the CMB given the best-fit 2015 \textit{Planck} \texttt{plikHM\_TT\_lowTEB\_lensing} lensed power spectra \citep{planck15-15}. A small expected galactic and extra-galactic Gaussian foreground contribution is also added, and then a smoothed version of the SPTpol beam window function is convolved. Note that because the TOD filtering is linear by construction and approximately diagonal in $QU$ Fourier space, it is not crucial that these simulations exactly match the true sky power, nor that they contain the right level of lensing or foreground non-Gaussianity.

From these, we generate mock TOD by virtually scanning the sky using the recorded pointing information from actual observations. For each scan strategy (\dsonehundred, \dsfivehundredLT, and \dsfivehundredfull), we mock-observe the simulated sky into TOD, process TOD into maps, and then coadd these maps in the same way as the real data. The first of the two improvements mentioned above is that we only use a subset of the actual observations (in practice, 20), since many observations have identical scan strategies and would have effectively identical transfer functions. In parallel to these full pipeline simulations, we also perform a simple projection of the beam-convolved CMB+foregrounds to the flat-sky, with no other filtering applied. 

We can achieve sufficient accuracy on $\op{T}$ with only 20 simulations by using a new variance canceling technique. This method computes the transfer function as, 
\begin{align}
  \op{T} = {\rm Diagonal}\left\langle {\rm Re} \left[ \frac{ \big (\Mpix\,f_{\rm full-pipeline}\big )_{QU,\mathbf{l}}}{ \big (\Mpix\,f_{\rm projected}\big )_{QU,\mathbf{l}}} \right] \right\rangle_{\rm 20\;\rm sims}
  \label{eq:tf}
\end{align}
where the $f$ in the numerator and denominator are the mock-observed and projected maps, respectively, and $\Mpix$ is the pixel mask. The presence of the projected map in the denominator cancels sample variance in the estimate leading to much quicker Monte Carlo convergence. However, this comes at the cost that Eq.~\eqref{eq:tf} is actually a biased estimate of the true effective transfer function. 

With a simple test, we can verify 1) that this bias is small, 2) that our approximation that $\op{T}$ is diagonal in $QU$ Fourier space is sufficient, 3) that there is negligible Monte Carlo error due to using only 20 pipeline simulations, and 4) that our usage of only 20 observations per simulation is valid. For a set of simulations separate than those used to estimate $\op{T}$ and using a different set of 20 observations within each simulation, we compare the result of the full pipeline simulation versus simply applying $\op{T}$ to the projected map for the same realization. In the top panel of Fig.~\ref{fig:validate_TF}, we show these difference maps, and in the bottom panel we show their power spectrum averaged over a few realizations. In both top and bottom panels, we multiply by the full mask, $\op{M}$, so as to pick out only modes relevant for the analysis. We see that the difference is 1--4 orders of magnitude below the noise spectrum, hence $\op{T}$ is a very accurate representation of the true transfer function, particularly at smaller scales which drive the lensing constraint. The final estimate of $\op{T}$ used in the analysis is shown in the bottom left panel of Fig.~\ref{fig:operators}. 

We note that the variance canceling technique employed here may be of wider use, but only if full pipeline simulations are not required to quantify uncertainty, otherwise a larger set of simulations is needed anyway. Here we did not need such a larger set because the Bayesian pipeline does not use simulations to quantify uncertainty at all, and because for the QE pipeline we have used simulations from the forward data model, as this model is demonstrated sufficiently accurate for our purposes.

\subsection{Noise covariance}
\label{sec:noisecov}

\begin{figure}
  \includegraphics[width=\columnwidth]{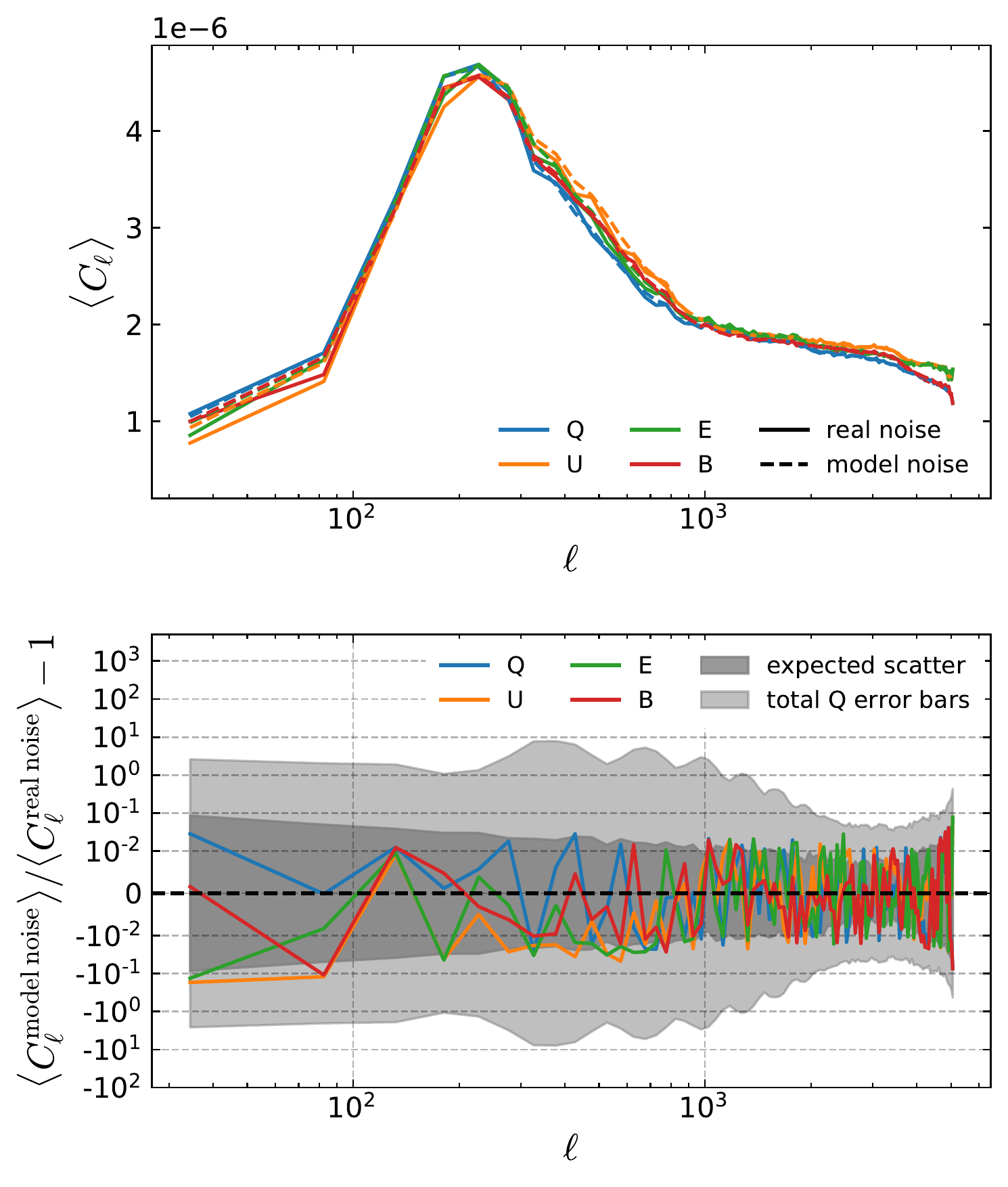}
  \caption{Validation of the approximations underlying our
  estimate of the noise covariance, $\Cn$ (see Sec.~\ref{sec:noisecov}). The top panel shows the mean power spectra of 400 real noise realizations and $10^4$ model noise realizations which have been masked by $\op{M}$. The bottom is a fractional difference between the two (note the change from linear to log scaling at $10^{-2}$). The dark shaded band is the expected scatter due to having only 400 real noise realizations, and the lighter shaded band gives the total CMB + noise error bars in the bins plotted here. The good agreement between the two indicates our model noise covariance is an accurate representation of the real noise.}
  \label{fig:real_vs_model_noise}
\end{figure}

The noise covariance is inferred from noise realizations which come directly from the real data using the ``sign-flipping'' method also used by previous SPT and BICEP analyses \citep[e.g.][]{bicep2collaboration2014,wu2019}. This method works by multiplying a random half of the $N\,{=}\,10490$ observations which enter the final data co-add by $-1$ before summing them. This cancels the signal but leaves the statistical properties of the noise unchanged, as long as no observation-to-observation correlations exist (which is expected to be the case). This is repeated $M\,{=}\,400$ times yielding $M$ nearly independent noise realizations. We will refer to these as {\it real noise realizations} and the distribution from which they are drawn as the {\it real noise}.

As we will describe in Sec.~\ref{sec:QE}, the QE pipeline only requires the average 2D power spectrum of the noise as well as an approximate white-noise level. This is sufficient because the noise only enters the QE pipeline for the purposes of Wiener filtering the data, where an approximate Wiener filter is computed and the impact of this approximation is captured in a Monte-Carlo correction applied at the end of the pipeline. This does not lead to any bias, only a small sub-optimality of the final result. The Bayesian pipeline does not apply any Monte-Carlo corrections, and thus needs to perform the Wiener filter (which also arises in the Bayesian case) more exactly. This in turn necessitates a full model for a noise covariance operator, $\Cn$, which needs to be as accurate as possible. We will refer to this as the {\it model noise}, and samples from this covariance as {\it model noise realizations}.

The real SPT noise is non-white, as instrumental and atmospheric $1/f$ noise dominates at large scales. It is anisotropic, as spatial modes in the scan-parallel and scan-perpendicular directions map onto different temporal modes, and are affected differently by TOD filtering. Finally, it is inhomogenous, as some spatial regions are observed slightly deeper than others; in particular, the lead-trail scanning strategy used in the \dsonehundred and \dsfivehundredLT observations causes some regions near the center and right edges of the final \dsultradeep field to have noise levels a few tens of percent lower than the rest of the field.

With only $M\,{=}\,400$ real noise realizations, but the most generic $\Cn$ corresponding to an $N_{\rm pix}\,{\times}\,N_{\rm pix}$ matrix where $N_{\rm pix}\,{=}\,2\,{\cdot}\,260^2$, some form of regularization is needed to choose a unique $\Cn$. The choice we make here is motivated by retaining the flexibility to model the complexity of the real noise just described while keeping $\Cn$ fast to invert and to square-root\footnote{We note that for our purposes, the matrix square-root is any $\op{G}$ for which $\Cn\,{=}\,\op{G}^\dagger \op{G}$.}, as both are needed to sample Eq.~\eqref{eq:jointposterior}. Specifically, we define the model noise covariance, $\Cn$, as
\begin{align}
  \Cn \equiv \op{W}\,\op{N}\,\op{W}^\dagger
\end{align}
where $\op{W}$ is diagonal in $QU$ pixel space and $\op{N}$ is diagonal in $QU$ Fourier space. That is to say, we model the noise as having an arbitrary non-white anisotropic power spectrum which is spatially modulated in pixel space. With this choice, we have that
\begin{align}
  \Cn^{-1} &= \op{W}^{-\dagger}\,\op{N}^{-1}\,\op{W}^{-1} \\
  \sqrt{\Cn} &= \sqrt{\op{N}}\,\op{W}^{\dagger},
\end{align}
where both operators can be easily applied to vectors with only a few FFTs. We solve for $\op{W}$ and $\op{N}$ by requiring that the variance in each individual 2D Fourier mode and the variance in each individual pixel be identical for noise realizations drawn from $\Cn$ and for the real noise realizations. These are $2N_{\rm pix}$ constraints for the $2N_{\rm pix}$ combined degrees of freedom in $\op{W}$ and $\op{N}$, yielding the following solution for the diagonal entries of these matrices
\begin{align}
  \op{W} &= {\rm Diagonal}\!\,\Big({\rm std}\left(\{n\}\right)_{QU,\mathbf{x}} \Big) \\
  \op{N} &= {\rm Diagonal}\!\,\Big({\rm var}\left(\{\op{M}\,\op{W}^{-1}n\}\right)_{QU,\mathbf{l}}\Big)
\end{align}
where the standard deviation and variance are taken across the $M$ noise realizations. 

We note that the noise realizations used in these averages are the raw sign-flipped combinations of the actual data, with no extra operators deconvolved or masks applied. Hence, the noise term, $n$, is not multiplied by any extra factors in Eq.~\eqref{eq:datamodel}. Additionally, we smooth both $\op{W}$ and $\op{N}$ with small Gaussian kernels, since we do not expect the noise properties to vary significantly across neighboring pixels or across neighboring Fourier modes. 

We plot $\op{W}$ and $\op{N}$ in the middle two panels of Fig.~\ref{fig:operators}. The top panel shows the spatially varying pixel variance pattern in $\op{W}$, and the bottom panel shows the non-white anisotropic Fourier noise pattern. To verify that model noise realizations drawn from $\Cn$ are largely indistinguishable from real noise realizations, we show in Fig.~\ref{fig:real_vs_model_noise} the mean $Q$, $U$, $E$, and $B$ power spectra of the 400 real noise realizations along with the mean power spectra of $10^4$ model noise realizations. We find excellent agreement, the difference between the two completely explained by the scatter expected due to having only 400 real noise realizations (dark shaded band). Additionally, any systematic difference between them is less than 1\% of the total Q sample variance error bars (lighter shaded band; note the switch from linear to log scaling at $10^{-2}$). As a further check, in Sec.~\ref{sec:simresults} we will use the model noise covariance to analyze simulated data which includes real noise realizations, finding no evidence for biases to $\Aphi$ due to any difference between these two.

\subsection{The noise-fill procedure}
\label{sec:noisefill}

The fact that $\Cn$ is not diagonal in either Fourier or map bases presents a challenge for exactly Wiener filtering the data in the presence of a masking operation which is also not diagonal in either space. Whether explicitly stated or not, computing such Wiener filters usually involves approximating the noise as diagonal in one of the two bases. Instead, here we develop and present the following procedure which can perform the operation exactly. To our knowledge, this has not been described before, and could be of general use. 

The challenge can be understood by considering the following toy problem. Suppose we observe some map which is the sum of some signal $s$ and noise $n$, both defined on the full pixel/Fourier plane, then apply a mask, $\op{M}$, which is a rectangular matrix mapping the full set of pixels/Fourier modes to a smaller subset of just the unmasked ones. The data model is thus given by $d=\op{M}(s+n)$. The residual between data and signal model is $(d - \op{M}\,s)$, and the covariance of this quantity is $\op{M}\, \op{N}\, \op{M}^\dagger$, where $\op{N}$ is the noise covariance. Defining the signal covariance as $\op{S}$, the log-posterior for this problem is thus
\begin{align}
  \log\mathcal{P}(s\,|\,d) &\propto -\frac{(d - \op{M}\,s)^2}{2\,\op{M}\, \op{N}\, \op{M}^\dagger} - \frac{s^2}{2\,\op{S}}.
\end{align}
Evaluating the posterior or its gradients with respect to $s$ requires inverting $\op{M}\, \op{N}\, \op{M}^\dagger$. Maximizing the posterior (i.e., Wiener filtering) requires this as well, as the solution is given by 
\begin{align}
  \hat s = \big[ \op{S}^{-1} + \op{M}^\dagger (\op{M}\, \op{N}\, \op{M}^\dagger)^{-1} \op{M} \big]^{-1} \op{M}^\dagger (\op{M}\, \op{N}\, \op{M}^\dagger)^{-1} d. 
\end{align}
However, since $\op{M}$ is not a square matrix, these inverses cannot be simplified away or trivially computed. Sometimes, as a simplifying assumption, $\op{M}$ and $\op{N}$ are taken to be diagonal in the same basis (e.g., $\op{N}$ is assumed to be white noise). In this case, the inverse can be computed explicitly (often in practice by setting the noise to infinity or to a very large floating point number). Since in our case we wish to not make this simplification, we cannot take this approach. 

The more general solution we use instead involves artificially filling in the masked data with extra noise, $\bar n$, such that the new data model is
\begin{align}
  d^\prime = d + \bar n = \op{M}(s+n) + \bar n,
\end{align}
where we are now considering $\op{M}$ as a square operator but with some rows which are zero. Note that the extra noise does not shift the mean of the data. However, the covariance of the data residual becomes
\begin{align}
  \op{M}\, \op{N}\, \op{M}^\dagger + \op{\bar N},
\end{align}
where $\op{\bar N}$ denotes the covariance for $\bar n$. Since we are free to choose $\op{\bar N}$, we can choose it such that the new data residual covariance is easy to invert, in particular, such that it is equal to $\op{N}$. This happens when
\begin{align}
  \op{\bar N} = \op{N} - \op{M}\, \op{N}\, \op{M}^\dagger.
\end{align}
We can draw a sample from $\mathcal{N}(0,\op{\bar N})$ by computing $\op{\bar N}^{\nicefrac{1}{2}} \xi$ where $\xi$ is a unit random normal vector. This can in turn be computed by evolving the following ordinary differential equation (ODE) from $t\,{=}\,0$ to $t\,{=}\,1$,
\begin{align}
  \label{eq:sqrtode}
  \frac{dy}{dt} = -\frac{1}{2}\big[\op{\bar N} t+(1-t) \mathbbgreek{1}\big]^{-1}(\mathbbgreek{1}-\op{\bar N}) y(t)
\end{align}
starting from $y(0)\,{=}\,\xi$ \citep{allen2000}. The quantity in brackets in Eq.~\eqref{eq:sqrtode} can be inverted with the conjugate gradient method. The ODE itself is requires a stiff solver (we use \texttt{CVODE\_BDF} from the \texttt{Sundials.jl} package; \citealt{hindmarsh2005,rackauckas2017}). The computation is not particularly costly and only needs to be done once at the beginning of any analysis. Once $d^\prime$ is computed, the new posterior is given by the much simpler
\begin{align}
  \log\mathcal{P}(s\,|\,d^\prime) &\propto -\frac{(d^\prime - \op{M}\,s)^2}{2 \, \op{N}} - \frac{s^2}{2\,\op{S}}.
\end{align}

\begin{figure}
  \includegraphics[width=\columnwidth]{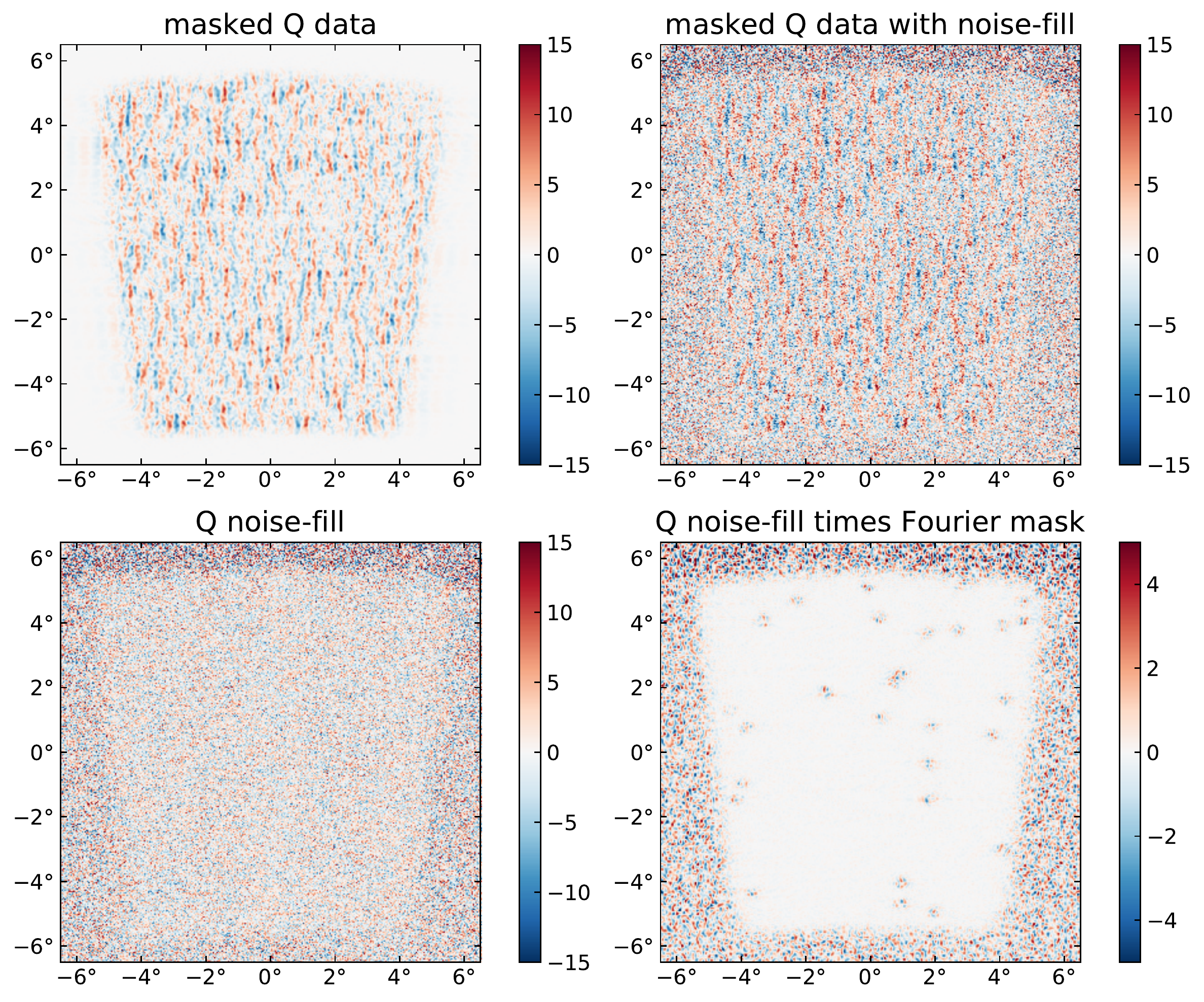}
  \caption{A demonstration of the ``noise-fill'' procedure described in \secref~\ref{sec:noisefill}, which makes it much easier to exactly Wiener filter the data even in the presence of pixel and Fourier space masking and a noise covariance model which is not diagonal in either space. The top-left panel shows \dsultradeep data with the mask applied, including Fourier and pixel masks. The top-right panel additionally has the noise-fill, $\bar n$, added in; this panel is exactly the data, $d$, which is used in the posterior in Eq.~\eqref{eq:jointposterior}. The bottom-left panel shows just $\bar n$, and the bottom-right panel is $\bar n$ multiplied by the Fourier mask. In this last panel, one can see that in the region interior to the mask and in the range of Fourier modes which are not masked by the Fourier mask, no extra noise is added (hence this procedure does not degrade constraints). Here we have plotted just the $Q$-polarization component; $U$-polarization behaves qualitatively the same.}
  \label{fig:noisefill}
\end{figure}

Note that, when generating simulated data, it is not necessary to actually perform this procedure. Instead, it is equivalent to simply generate data from a model $d = \op{M}\,s+n$, i.e., to leave the noise unmasked. It is only on the real data, where one does not have access to $s$ and $n$ separately, that one needs to explicitly perform the noise-fill. An added benefit of this approach is that the likelihood term in the posterior becomes a full $N_{\rm pix}$-dimensional $\chi^2$, thus its expectation value and scatter are easy to compute; we use this in the later sections to ascertain goodness-of-fit. Finally, note that $\bar n$ is generally zero ``inside'' the mask and only non-zero ``outside'' the mask, thus the degradation in constraints due to the filled in noise is negligible. We have verified this by filling in our real data with several different noise realizations, finding no shift in the resulting constraints on $\Aphi$. In \figref~\ref{fig:noisefill}, we plot example data and noise-fills for the \dsultradeep dataset. 

\subsection{Negligible effects}

To conclude this section, we mention a few effects which are expected to be negligible for this data set and are thus not modeled. Both Bayesian and QE pipelines ignore sky curvature, instead working in the flat-sky approximation, which is very accurate for the modestly sized 100\,deg$^2$ patch considered here. The lensing operation is implemented with \LenseFlow \citepalias{millea2020}, which assumes the Born approximation. Post-Born effects are not detectable until much lower noise levels and are thus ignored \citep{pratten2016,fabbian2018,bohm2018,beck2018}. Finally, we do not model galactic or extra-galactic foregrounds. The \dsultradeep field is in a region of sky particularly free of galactic contamination, and we conservatively mask modes below $\ell\,{\sim}\,500$, thus we expect negligible polarized galactic dust foregrounds \citep{planckcollaboration2020}. Extra-galactic foregrounds are expected to be much smaller in polarization than in temperature, and here we only use polarization. Given that we also conservatively mask modes above $\ell\,{=}\,3000$, we follow \cite{wu2019} in concluding extra-galactic foregrounds can be ignored in this analysis.

\section{Lensing analysis}
\label{sec:lensing}

\subsection{Bayesian lensing}
The Bayesian sampling pipeline very closely follows the methodology described in \citetalias{millea2020}, and uses the same code, \textsc{CMBLensing.jl}\,\href{https://github.com/marius311/CMBLensing.jl}{\faGithub}. Conceptually it is extremely straight-forward: it is simply a Monte-Carlo sampler of the full posterior given in Eq.~\eqref{eq:jointposterior}. Beyond this, there are a few practical details which we describe in this section. 

First, we perform the standard change-of-variables from $(f,\phi)\rightarrow(f^\prime,\phi^\prime)$ and sample the posterior in terms of $(f^\prime,\phi^\prime)$ instead. In this parameterization, the posterior is less degenerate and better conditioned, yielding much better performance of the sampling algorithm. This was extensively discussed in \citetalias{millea2020}, and we apply the same reparametrization as described there almost without change. Specifically, we take 
\begin{align}
  \label{eq: def phimix} \phi^\mix &\equiv \opG[\Aphi]\, \phi \\
  \label{eq: def fmix} f^\mix      &\equiv \Len[\phi] \, \D \, f .
  \end{align}
The operator $\D$ is defined to be diagonal in $EB$ Fourier space, and $\opG[\Aphi]$ is diagonal in Fourier space, with
\begin{align}
  \label{eqn: Dr}
  \D &\equiv \left[ \frac{\Cf + 2 \, \op{N}_f}{\Cf} \right]^{\nicefrac{1}{2}} \\
  \label{eqn: GA}
  \opG[\Aphi] &\equiv \left[ \frac{\Cphi[\Aphi] + 2 \, \op{N}_\phi}{\Cphi[\Aphi]} \right]^{\nicefrac{1}{2}}
\end{align}
where $\op{N}_f$ should approximate the sum of instrumental noise and lensing-induced excess CMB power, and $\op{N}_\phi$ should approximate noise in the $\phi$ reconstruction. Here, we find a sufficient choice is to set $\op{N}_f$ to isotropic 12\,$\mu$K-arcmin white noise, and $\op{N}_\phi$ to the 2D QE $N^{(0)}$ bias. We note that the optimal choice of these operators is not precisely defined, and poor choices do not affect results, instead only lead to slower convergence.

With the reparametrized target posterior in hand, we now describe the sampler. For both convenience and efficiency, the sampling is broken up into separate Gibbs steps where we sample different conditional slices of Eq.~\eqref{eq:jointposterior}. The Gibbs procedure ensures that after a sufficiently long time, the chain of conditional samples asymptotes to draws from the joint distribution. 

The first Gibbs step samples the conditional distribution of $f$ given the other variables. The advantage of splitting this off as its own Gibbs step is that this conditional is Gaussian and can be sampled exactly by running one conjugate gradient solver. This solver involves inverting the operator shown below in Eq.~\eqref{precon1}, where we have left out instrumental parameters and beam and transfer functions for clarity.\footnote{The exact operator to be inverted can be derived by taking the derivative $d/df$ of Eq.~\eqref{eq:jointposterior}, setting it equal to zero, and solving for $f$.} We use a nested preconditioner wherein we precondition Eq.~\eqref{precon1} with Eq.~\eqref{precon2}, which itself involves a conjugate gradient solution using Eq.~\eqref{precon3} as a preconditioner. In Eq.~\eqref{precon3} we use a noise operator, $\op{\widehat C}_{n}$, which is an approximate $EB$ Fourier-diagonal version of $\Cn$, making the final preconditioner explicitly invertible.
\begin{alignat}{3}
  &\Cf^{-1} \;\; +& \;\; \Len[\phi]^\dagger \, \Mpix^\dagger \, \Mfourier^\dagger \, &\Cn^{-1} \, \Mfourier \, \Mpix \, \Len[\phi] \label{precon1} \\
  &\Cf^{-1} \;\; +& \;\; \Mpix^\dagger \, \Mfourier^\dagger \, &\Cn^{-1} \, \Mfourier \, \Mpix \label{precon2} \\
  &\Cf^{-1} \;\; +& \;\; \Mfourier^\dagger \, &\op{\widehat C}_{n}^{-1} \, \Mfourier \label{precon3}
\end{alignat}
The advantage of this scheme is that it minimizes the number of times we need to compute the action of Eq.~\eqref{precon1}, which involves two lensing operations and hence is much costlier than the others. With the nested preconditioning, only a few applications of Eq.~\eqref{precon1} are necessary per solution.

The second Gibbs step samples the conditional distribution of $\phi$ given the other variables. This sample is drawn via Hamiltonian Monte Carlo \citep{betancourt2017}, which involves sampling a random momentum, $p_\phi$, from a chosen mass matrix, and then performing a symplectic integration to evolve the Hamiltonian for the system. Poor choices of mass matrix or large symplectic integration errors yield a slower converging chain, but do not bias the result asymptotically. We find that 25 leap-frog symplectic integration steps with step size $\epsilon\,{=}\,0.02$ per Gibbs pass yield nearly optimal convergence efficiency. We note that to control symplectic integration error, we also need at least a 10-step 4th-order Runge-Kutta ODE integration as part of the \LenseFlow solver (in \citetalias{millea2020}, only 7 steps were needed, likely due to simpler masking). Finally, the mass matrix should ideally approximate the Hessian of the log-posterior; here we use, 
\begin{align}
  \label{eq:hessphi}
  \mathbbgreek{\Lambda}_{\phi}(\Aphi) = \opG[\Aphi]^{-2} \Big[\,\mathbb N_\phi^{-1} +  \Cphi[\Aphi]^{-1}  \Big]
\end{align}

The final Gibbs passes sample the conditionals of each of the remaining scalar parameters in turn: $\Aphi$, $\Pcal$, $\psipol$, $\epsQ$, $\epsU$, and the $\beta_i$. Since these are one-dimensional distributions, we sample by evaluating the log-posterior along a grid of values, interpolating it, then using inverse-transform sampling to get an exact sample. Importantly, in all cases except $\Aphi$, these parameters are "fast" parameters because $\Len[\phi]f$ remains constant along the conditional slice and can be computed just once at the beginning of the pass. Indeed, sampling these parameters accounts for $<5\%$ of the total runtime of a chain, and one could imagine adding many other instrumental parameters like these at almost no computational cost. Sampling $\Aphi$ is somewhat costlier because Eq.~\eqref{eq: def phimix} couples $\Aphi$ and $\phi$, meaning that each grid point of $\Aphi$ requires lensing a new map (however, the decorrelating effect of the reparametrization far outweighs this increased computational cost).

\subsection{Quadratic estimate}
\label{sec:QE}

The QE analysis closely follows those of the \SI{100}{deg \squared} and \SI{500}{deg \squared} SPTpol analyses \citep{story15,wu19}. It uses the standard SPT QE pipeline, and so is completely independent from the Bayesian code. We give a brief review of the QE pipeline here and take note of aspects particular to this analysis, referring the reader to the previous works for a more comprehensive treatment.

The QE uses correlations between Fourier modes in pairs of CMB maps to estimate the lensing potential; here we use the same modified form of the \citet{hu02} estimator as in \citet{wu19},
\begin{align}
    \bar \phi_{\vb{L}}^{XY}  &= \int \dd[2]{\vb*{\ell}} \bar X_{\vb*{\ell}} \bar Y_{\vb*{\ell} - \vb{L}}^*  W_{\vb*{\ell}, \vb*{\ell} - \vb{L}}^{XY},
    \label{eq:QE}
    \end{align}
where $\bar X$ and $\bar Y$ are inverse-variance filtered data maps and $W^{XY}$ is a weighting function with $XY\,{\in}$\,\{EE, EB\}. 

The inverse-variance filtering used for the QE does not employ the noise-fill procedure outlined in Sec.~\ref{sec:noisefill}, opting instead to leave the existing pipeline unmodified. Here, the noise is approximated as the sum of two components. The first is a pixel-space diagonal component, $\op C_{n,p}\,{=}\,\Mpix^{-1} \, \op Z \, \Mpix^{-1}$, where $\Mpix$ is the pixel mask and $\op Z$ is a homogeneous white noise covariance specified by the noise levels at the end of Sec.~\ref{sec:data}. The second is a Fourier-space diagonal component, $\op C_{n,f}$, which includes the power spectrum of atmospheric foregrounds and excess instrumental $1/f$ noise not captured in the first component, and is determined empirically from the real noise realizations. Inverse variance filtering can then be performed by solving the following equation for $\bar X$ with conjugate gradient:
\begin{align}
    \qty [\op S^{-1} + \op F^{\dagger} \, \op C_{n,p}^{-1} \, \op F] \, \op S \, \bar{X} &= \op F^{\dagger} \, \op C_{n,p}^{-1} \, d_{\rm QE},
\end{align}
where $\op S = \Cf + \op C_{n, f}$ and $\op F = \op T \, \op B$.

We then correct each estimator, $\bar \phi_{\vb{L}}^{\,XY}$, by 1) subtracting a mean-field bias, $\bar\phi_{\vb{L}}^{XY, \rm MF}$, computed from an average over simulations, 2) normalizing by the analytic response, $R_{{\vb{L}}}^{XY, \rm Analytic}$, and 3) summing the debiased and normalized estimates. We account for the impact of the pixel mask, not captured by the analytic response, with an isotropic Monte Carlo correction, $R_L^{\rm MC}$. This is computed by fitting a smooth curve to the ratio $C_{\ell}^{\bar \phi \times \phi_{\rm true}} / C_{\ell}^{\phi\phi, \rm theory}$, averaged over simulations. This gives a normalized unbiased estimate
\begin{align}
    \hat \phi_{\vb{L}} = \frac{1}{R_{L}^{\rm MC} } \frac{\sum_{XY} \; \bar\phi_{\vb{L}}^{XY} - \bar\phi_{\vb{L}}^{XY, \rm MF} }{\sum_{XY} R_{\vb{L}}^{XY, \rm Analytic}}.
\end{align}

\begin{figure}
  \includegraphics[width=\columnwidth]{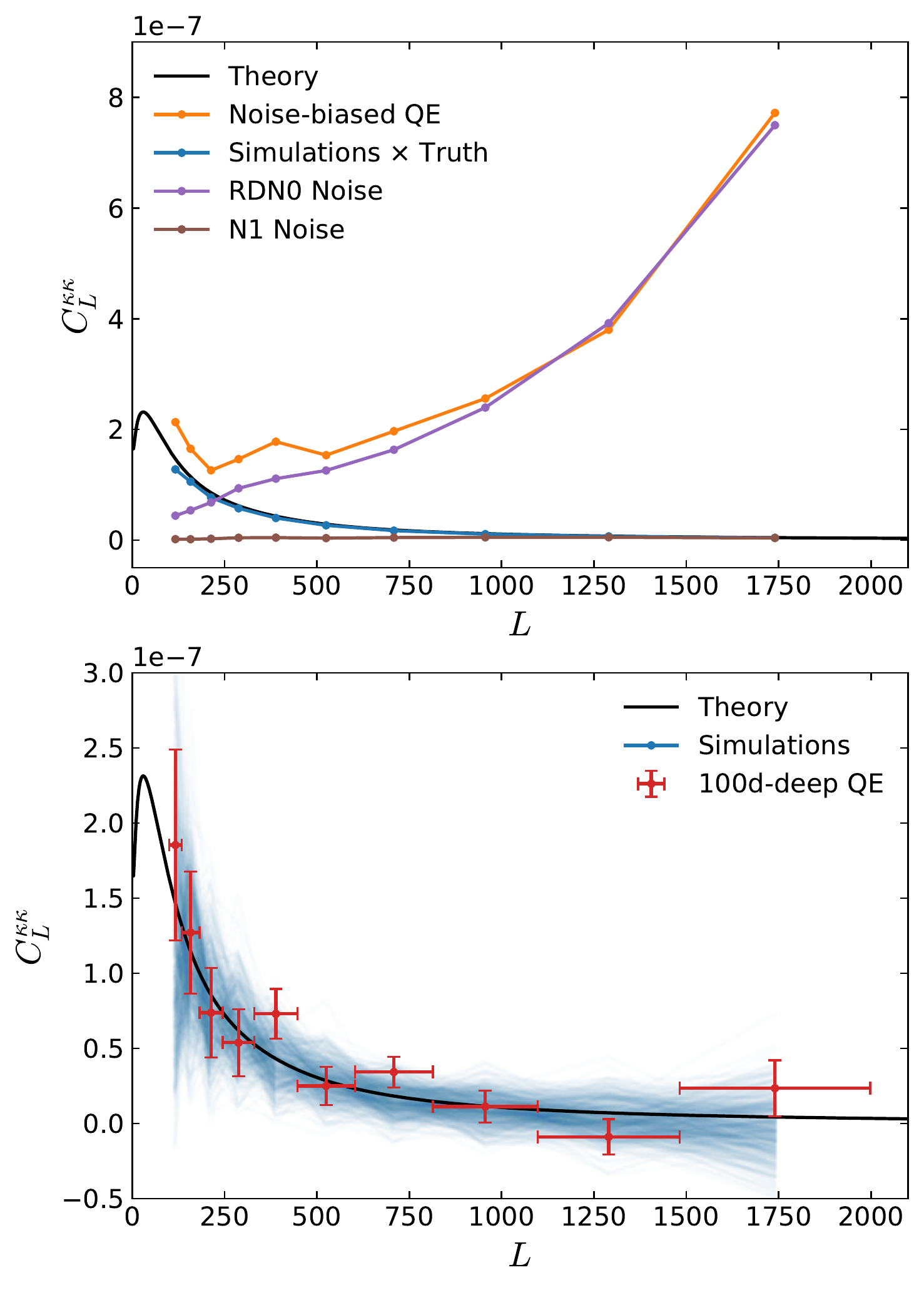}
  \caption{Bandpowers and noise terms from the quadratic estimate (QE) pipeline. The top panel shows the normalized but noise-biased QE power spectrum, along the typical $N^{(0),\rm RD}_L$ and $N^{(1)}_L$ noise biases which are subtracted. The blue curve is the average cross spectrum between input $\phi$ maps and $\bar \phi_{\vb{L}}^{XY}$ across a suite of simulations, and is used to compute $R_{L}^{\rm MC}$. The bottom panel shows the noise-bias-subtracted QE and error bars (from simulations), as well as a cloud of blue lines denoting the noise-debiased simulations used to compute $f_{\rm PS}$.}
  \label{fig:qe_spectra}
\end{figure}
  
To obtain constraints on $A_{\phi}$, we take the autospectrum of $\hat\phi_{\vb{L}}$ to form biased lensing power spectra, $\bar C_{\ell}^{\phi \phi}$. We then estimate the typical $N_L^{(0), \rm RD}$ and $N_L^{(1)}$ biases using simulations, and apply a final multiplicative MC correction $f_{\rm PS}$ as in \citet{wu19}. No foreground correction is applied, so the final expression for the debiased bandpowers is
\begin{align}
    \widehat C_{\ell}^{\phi \phi} = f_{\rm PS} \qty[\bar C_{\ell}^{\phi \phi} - N_L^{(0), \rm RD} - N_L^{(1)}].
\end{align}
We calculate the covariance between the bandpowers, $\Sigma$, by running a Monte Carlo over the entire procedure. Fig.~\ref{fig:qe_spectra} shows the bandpowers of $\widehat C_L^{\kappa \kappa}\,{\equiv}\,L^4 \widehat C_L^{\phi \phi}/4$, along with error bars computed from the diagonal of $\Sigma$.

Since the bandpower errors are assumed Gaussian, the resulting $\Aphi$ constraints are also Gaussian, and are given by
\begin{align}
  \hat A_\phi^{\rm QE} &= \frac{\widehat C_{\ell}^{\phi \phi} \, (\Sigma^{-1})_{\ell\ell^\prime} \, C_{\ell^\prime}^{\phi \phi}}{C_{\ell}^{\phi \phi} \, (\Sigma^{-1})_{\ell\ell^\prime} \, C_{\ell^\prime}^{\phi \phi}} \\
  \sigma(A_\phi^{\rm QE}) &= \frac{1}{\sqrt{C_{\ell}^{\phi \phi} \, (\Sigma^{-1})_{\ell\ell^\prime} \, C_{\ell^\prime}^{\phi \phi}}},
\end{align}
where the summation over $\ell$ is implied. For this calculation, we truncate $\Sigma$ at the third off-diagonal, beyond which we do not resolve any non-zero covariance to within Monte Carlo error, consistent with the expectation that the correlation should be small for very distant bins. We note, however, that correlation between neighboring bins can be as large as 10\% and has a significant impact on the final uncertainties.

\section{Validation}
\label{sec:validation}

\subsection{Chain convergence}

\begin{figure}
  \includegraphics[width=\columnwidth]{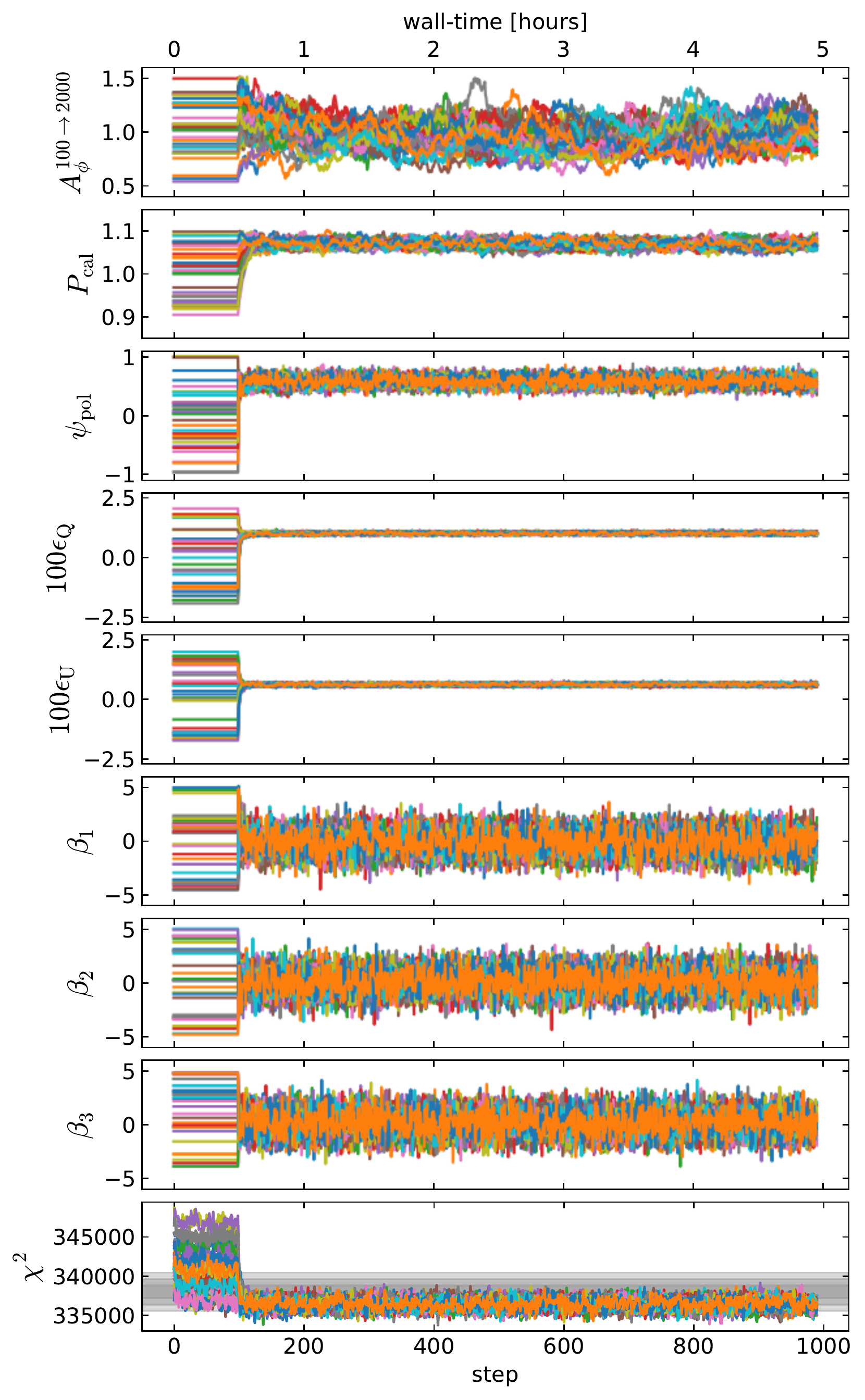}
  \caption{The top 8 plots show the trace of the sampled cosmological and systematics parameters, $\theta$, at each step in the Monte Carlo chain. The very bottom plot shows the trace of the $\chi^2$ of the current model point, along with a gray shaded band indicating the expectation based on the number of degrees of freedom. Note that 202,800 other parameters are jointly sampled in this chain (not pictured), corresponding to every pixel or Fourier mode in the CMB polarization and $\phi$ maps. To aid convergence, the $\theta$ are not updated for the first 100 steps in the chain. These 32 independent chains ran across 4 Tesla V100 GPUs in roughly 5 hours.}
  \label{fig:trace}
\end{figure}

\begin{figure*}
  \includegraphics[width=\textwidth]{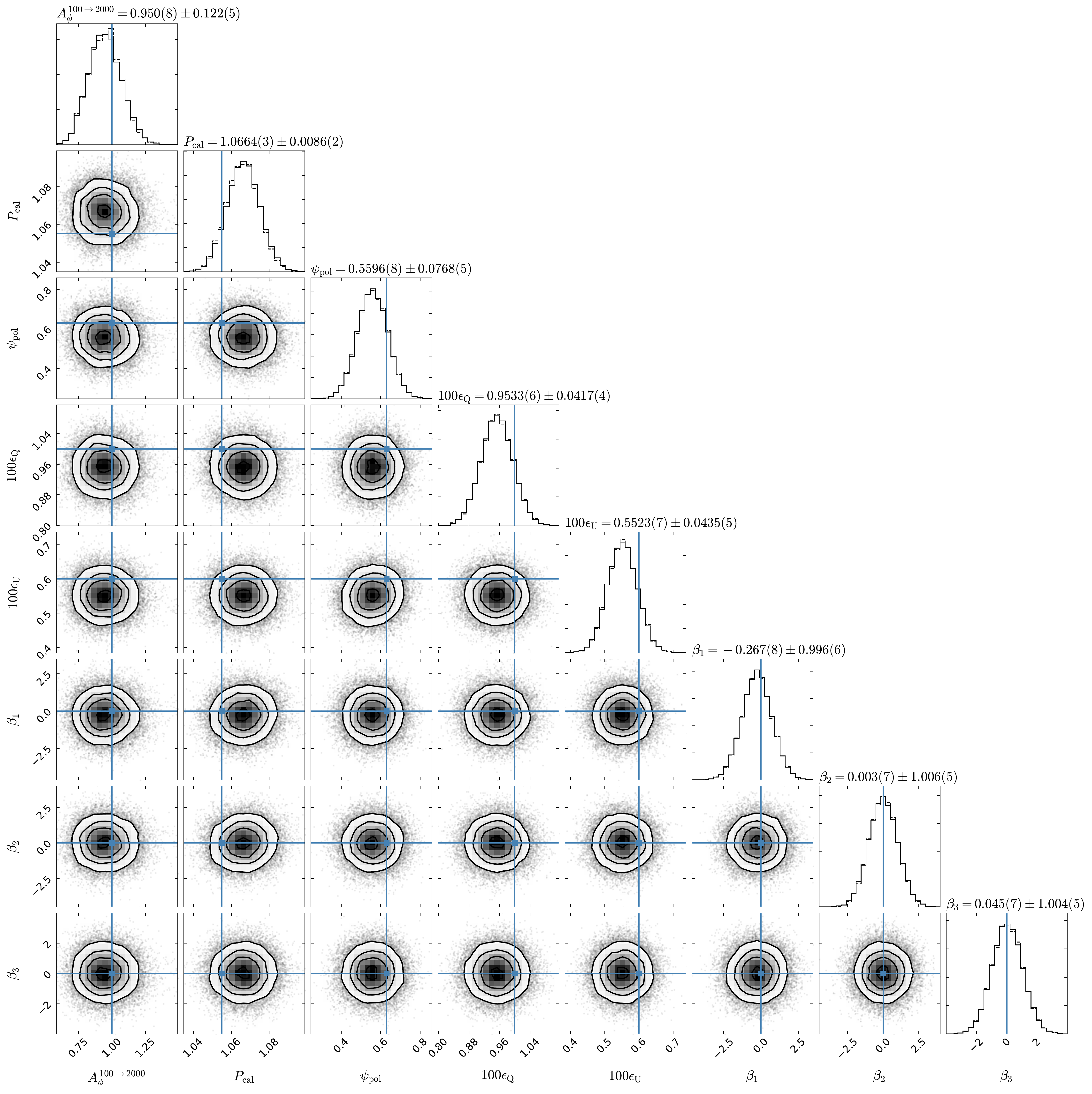}
  \caption{Constraints on sampled parameters, $\theta$, from our baseline \dsultradeep chain. The two-dimensional plots show 1, 2, and 3\,$\sigma$ posterior contours as black lines, with binned 2D histograms of the samples shown inside of the 3\,$\sigma$ boundary and individual samples shown beyond that. The first column is the main cosmological parameter of interest $\Aphiarrow$, and the remaining columns are systematics parameters. The ability to easily and jointly constrain cosmological and systematics parameters in this manner, while implicitly performing optimal lensing reconstruction and delensing, is a unique strength of the Bayesian procedure. Here, we find ${<}\,5\%$ correlation between $\Aphiarrow$ and any systematics, meaning $\sigma(\Aphiarrow)$ is increased by ${<}\,2\%$ upon marginalizing over systematic uncertainty. For the systematics parameters, the blue lines denote an estimate from an external procedure, and the agreement in all cases is an important consistency check. The 1D histograms also include the posterior from a separate independent chain as a dashed line, indicating the distributions are sufficiently well converged. More quantitatively, the numbers in parenthesis in the titles give an estimate of the standard error on the last digit of the posterior mean and of the posterior standard deviation.}
  \label{fig:baseline_triangle}
\end{figure*}

\begin{figure*}
  \includegraphics[width=\textwidth]{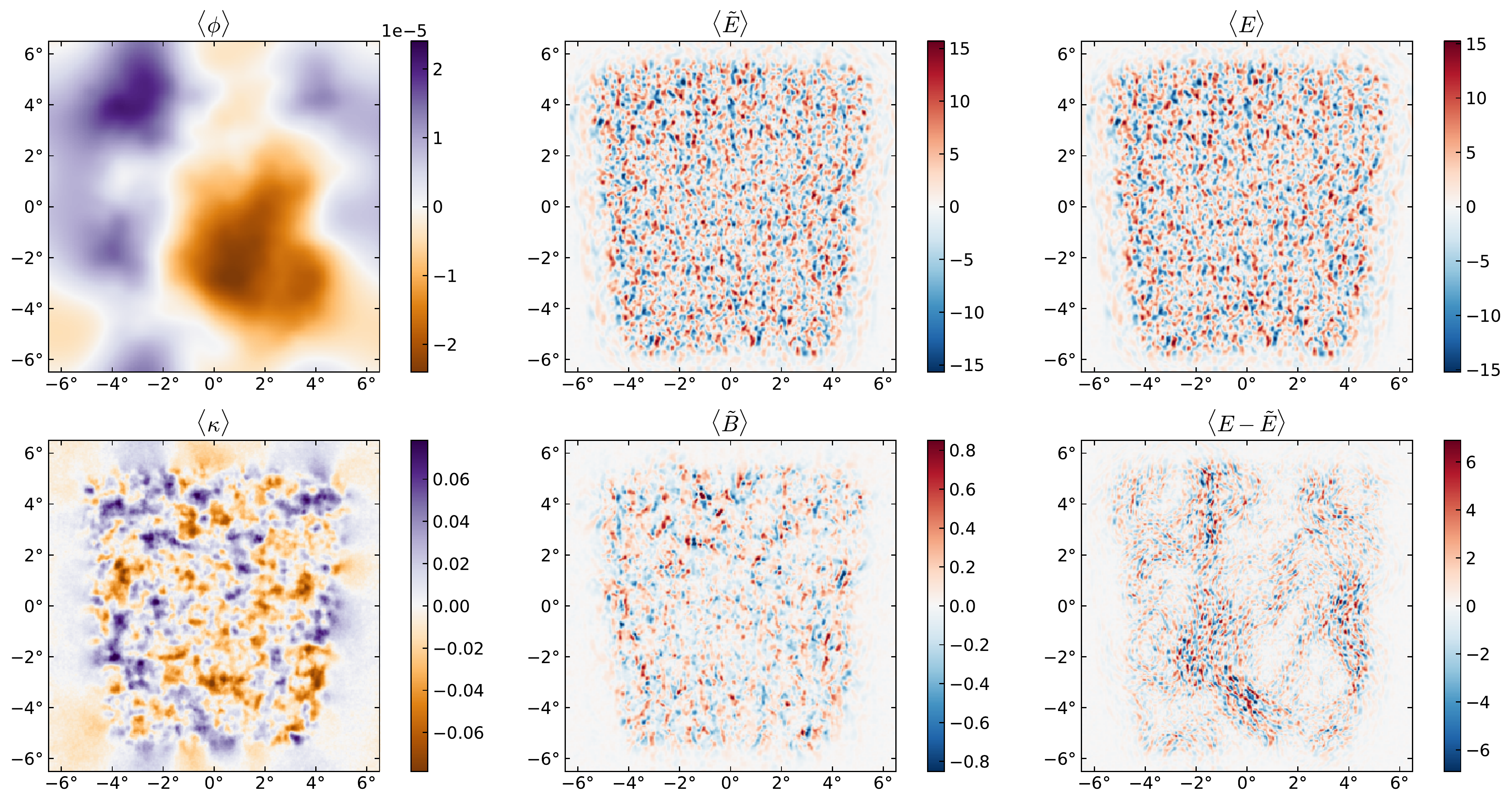}
  \caption{Posterior mean maps, computed by averaging over the Monte Carlo samples in our chains. The quantities $\phi$ and $\kappa\equiv-\nabla^2\phi/2$ are the lensing potential and convergence maps, and $\tilde E$ and $\tilde B$ are the lensed $E$ and $B$ mode polarization maps. The posterior of any quantity can be computed by post-processing the chain and averaging; for example, the bottom right panel shows the posterior mean of $(\tilde E - E$), i.e. the lensing contribution to the $E$ mode map. These maps are in some sense only a byproduct of the $\Aphi$ inference, but if a single point estimate of any of these quantities is required elsewhere, these are the best estimates to use. As expected, these maps qualitatively resemble Wiener filtered data, wherein low signal-to-noise modes are suppressed. The Monte Carlo error in these maps is more quantitatively explored in Fig.~\ref{fig:posterior_mean_bandpowers}.}
  \label{fig:posterior_mean_maps}
\end{figure*}

\begin{figure*}
  \includegraphics[width=\textwidth]{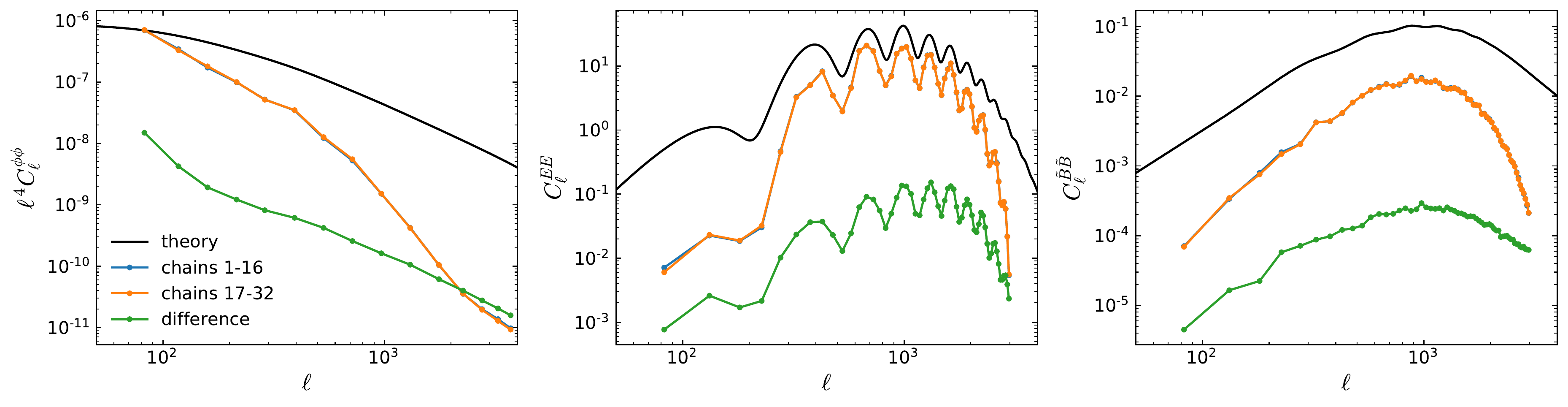}
  \caption{The blue and orange lines (nearly coincident) show the power spectra of (from left to right) posterior mean $\phi$, unlensed $E$, and lensed $B$ maps, as determined from one half of the 32 independent \dsultradeep MCMC chains vs. the other half. The power spectra of posterior mean maps is expected to be suppressed relative to theory, similar to the suppression that arises when Wiener filtering. The green line shows the power spectra of the map differences between these two sets of chains. Across almost all scales, these differences are 1--2 orders of magnitude below the spectra of the mean maps, demonstrating the level of convergence of these chains. The smallest scales in $\phi$ are the only region where the difference is larger than the mean. An analysis which required better accuracy here could run more chains, although we note these scales do not impact the determination of $\Aphiarrow$.}
  \label{fig:posterior_mean_bandpowers}
\end{figure*}

One of the main challenges of the Bayesian procedure is ensuring the Monte Carlo chains are sufficiently converged and are thus yielding stationary samples from the true posterior distribution. A large body of work exists on verifying chain convergence, and many methods of varying sophistication exist. Our experience has been that the most robust and accurate check is actually the simplest, namely just running multiple independent chains in parallel starting from different initial points, and ensuring that the quantities of interest have identical statistics between the different chains. Here, we are in a fortunate position where this is possible, largely because: 1) it is computationally feasible to run many chains and to run existing chains for longer if there is any doubt, and 2) we find no evidence for complicated multi-modal distributions, so convergence is not about finding multiple maxima but rather simply a matter of getting enough samples to smoothly map out the (mildly) non-Gaussian posteriors of interest.

Checking for convergence usually begins by visually inspecting the samples from a chain. For the baseline \dsultradeep chain, we show the sampled values of the cosmological and systematics parameters comprising $\theta$ in Fig.~\ref{fig:trace}. Our default runs evolve 32 chains in parallel (batches of 8 chains per Tesla V100 GPU) and hold $\theta$ fixed for the first 100 steps to give the $f$ and $\phi$ maps a chance to find the bulk of the posterior first, which reduces the needed burn-in time. Note that the starting point for our chains are a sample from the prior, not just for $\theta$ but also for the $\phi$ and $f$ maps themselves.\footnote{Note that due to the ``curse of dimensionality'', these random starting points are much further apart in the high-dimensional parameter space than might seem from looking at any 1D projection.} Despite this, Fig.~\ref{fig:trace} shows that all $\theta$ converge to the same regions in parameter space, and no ``long wavelength'' drift is seen in the samples.

We also check convergence by splitting the 32 chains into two sets of 16 and estimating parameter constraints from each set. The 1D posteriors from two sets of the baseline \dsultradeep case are shown in Fig.~\ref{fig:baseline_triangle}. Here we remove a burn-in period of 200 samples from the beginning of each chain. We find that all contours overlap closely and no conclusions would be reasonably changed by picking one half over the other.

To make the convergence diagnostics more quantitative, we use the following procedure throughout this paper whenever quoting any number derived from a Monte Carlo chain. We first compute the effective sample size (ESS) of the quantity of interest given the observed chain auto-correlation \citep{goodman2010}. We then use bootstrap resampling to estimate the Monte Carlo error, wherein 1) we draw $N$ random samples with replacement from the chain where $N$ is the ESS, 2) we compute the quantity in question using these samples, then 3) we repeat this thousands of times and measure the scatter. The scatter gives a $1\,\sigma$ Monte Carlo error which we report using the typical notation that $M$ digits in parentheses indicate an error in the last $M$ digits of the quantity, i.e. $1.23(4)$ is shorthand for $1.23\,{\pm}\,0.04$. We use this not only for the posterior mean, but also standard deviations, correlation coefficients, or any other quantity estimated from the chain.

For example, skipping ahead to the results presented in the next section, the constraint on $\Aphi$ from the \dsultradeep chain is
\begin{align}
  \Aphi = 0.949(8) \pm 0.122(5)
  \label{eq:ultradeep_Aphi}
\end{align}
This is to say, the standard error on the mean is 0.008, which is an acceptable 6\% of the 1\,$\sigma$ posterior uncertainty of 0.122(5), and could be reduced further by running the chain longer if desired. 

If we are interested only in constraints on $\Aphi$, then Eq.~\eqref{eq:ultradeep_Aphi} gives us what we need to know about how accurate our posterior inference on this quantity is. It is the case, however, that not all modes in the corresponding $\phi$ samples in the chain are necessarily converged to this same level. This will not affect $\Aphi$ since not all modes are informative for $\Aphi$, and the errors in Eq.~\eqref{eq:ultradeep_Aphi} tell us about the convergence of the sum total of all modes which {\it are} informative. In other applications, however, we might care about other modes, for example for delensing external datasets or for cross correlating with other tracers of large scale structure. We can check the convergence for all modes at the field level by computing posterior mean maps and comparing the power spectrum of the difference when estimated again from two independent sets of 16 chains. Fig.~\ref{fig:posterior_mean_maps} shows posterior mean maps and Fig.~\ref{fig:posterior_mean_bandpowers} shows the power spectrum differences from the two independent sets. Across a wide range of scales in $\phi$, $E$, and $B$, the power of the difference maps is 1--2 orders of magnitude below the signal. The only exception is very small scales in $\phi$; indeed, this is an example of modes for which the standard error is larger than the mean, but which are not informative for $\Aphi$. If one uses these samples for a downstream analysis, one could use the bootstrap resampling procedure with the maps themselves to estimate the Monte Carlo error in whatever final quantity was computed from these samples.

\begin{figure}
  \includegraphics[width=\columnwidth]{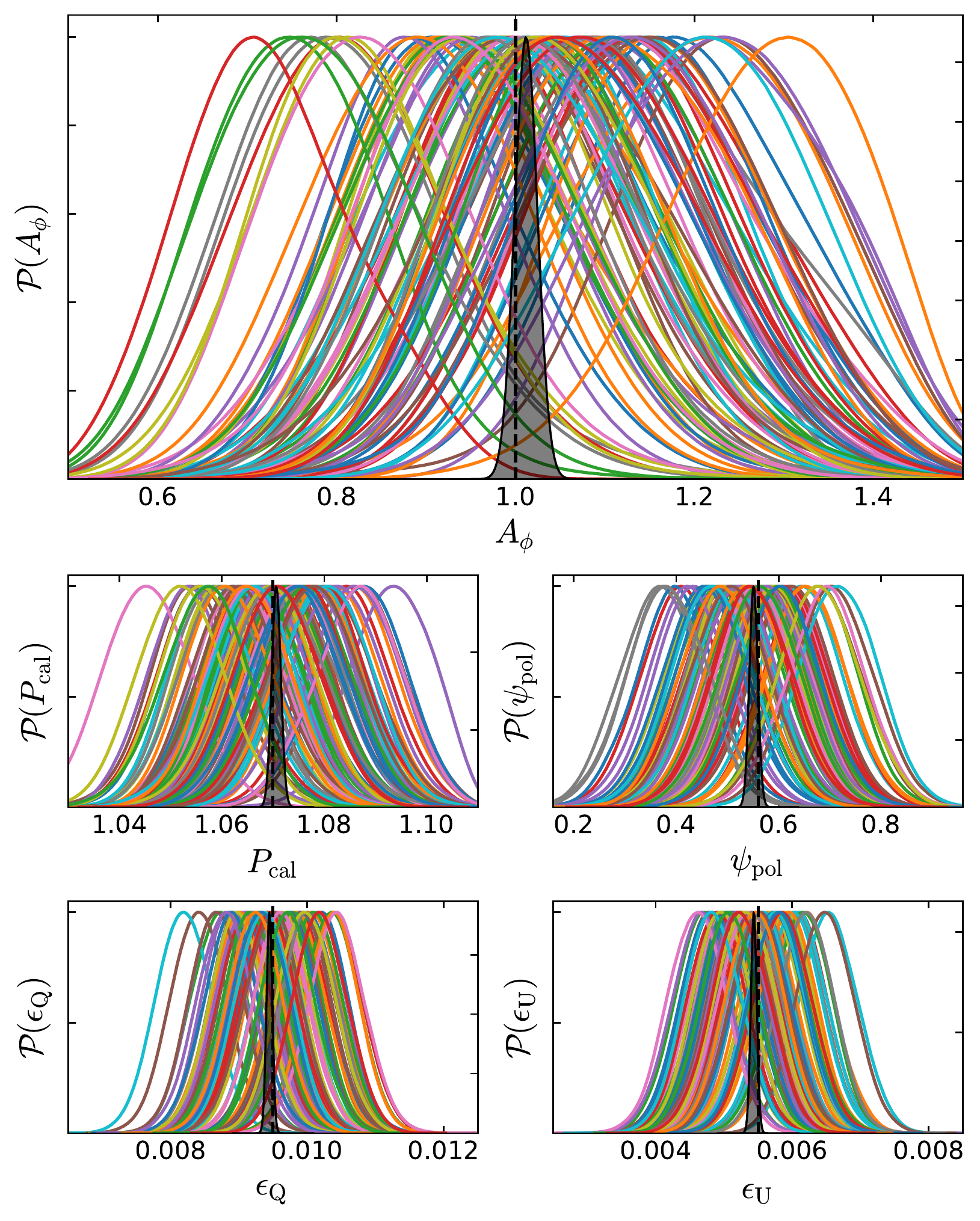}
  \caption{Validation of the Bayesian pipeline on simulations. Colored lines in each panel denote the posterior distributions from each of 100 simulated \dsultradeep data sets (these include real noise realizations). The shaded black curve is the product of all of these probability distributions. Note that, for clarity, all distributions have been normalized to their maximum value. The true value of the systematics parameters in these simulations comes from the best-fit \dsultradeep results, and are denoted by vertical dashed lines. The shaded black curve bounds possible systematic errors in the Bayesian pipeline due to mismodeling of the instrumental noise or pipeline errors, and we find no evidence for either to within the 10\% of the statistical error afforded by the 100 simulations.}
  \label{fig:sims}
\end{figure}

\vspace{1cm}
\subsection{Simulations}
\label{sec:simresults}

Having verified in the previous section that Monte Carlo errors in our chains are sufficiently small, we now verify the pipeline itself, as well as our noise covariance approximation. This is done by running chains on simulated data and checking that, on average, we recover the input truth. Crucially, the simulations we use include real noise realizations, while the posterior itself uses the model noise covariance. If the statistics of the real noise were different in a way not captured by the model noise covariance, we would expect to see some bias against the input truth in these simulations.

Fig.~\ref{fig:sims} shows these posterior distributions. The simulation truth uses the same fiducial {\it Planck} cosmology used in the baseline model (Sec.~\ref{sec:modeling}). Additionally, we include simulated systematics at a level given by the best-fit values of the \dsultradeep analysis itself, to confirm that we recover non-zero values of the systematics parameters. The colored lines are the posteriors from each of the $N\,{=}\,100$ simulations performed, and the shaded black curve is the product of all $N$. Because the simulated data are independent (ignoring the very small correlations between our sign-flipped noise realizations) and because the $\theta$ shown in this figure have a uniform prior, the product can also be interpreted as a single posterior given $N$ data, 
\begin{align}
  \mathcal{P}(\theta\,|d_1)\mathcal{P}(\theta\,|d_2)...\mathcal{P}(\theta\,|d_N) = \mathcal{P}(\theta\,|d_1,d_2,...,d_N)
\end{align}
This indicates that the black shaded contour should also, on average, cover the input truth. If there were any systematic biases affecting the inference of $\theta$, either from noise mis-modeling or from errors in the pipeline, we would expect to find a noticeable bias, which we do not. With $N{=}\,100$ simulations, we have formally checked against biases at the level of $1/\surd N\,{=}\,10\%$ of the $1\,\sigma$ error bar for any single realization.

\vspace{1cm}
\section{Results}
\label{sec:results}

\subsection{Joint $\Aphi$ and $\AL$ constraints}
\label{sec:AphiAL}

The $\Aphi$ constraint obtained from the QE explicitly does not use information from the power spectrum of the data because the weights $W_{\vb*{\ell}, \vb*{\ell} - \vb{L}}^{XY}$ in Eq.~\eqref{eq:QE} are zero when $\vb{L}\,{=}\,0$. The Bayesian constraint, however, extracts all information, including whatever may be contained in the power spectrum, as well as in all higher-order moments (bispectra, trispectra, etc...). To facilitate a more fair comparison between the two, and as a consistency check, it is useful to separate out the power spectrum information in the Bayesian case. 

A natural way to do so is by adding a correction to the noise covariance operator such that, 
\begin{align}
  \Cn \; \rightarrow \; \Cn + \DAL \, \op{A} \, \op{C}_{\rm len} \, \op{A}^\dagger,
  \label{eq:CnAL}
\end{align}
where $\DAL$ is a new free parameter, $\op{A}\,{\equiv}\,\op{M}\,\op{T}\,\op{B}$, and
\begin{multline}
  \op{C}_{\rm len} = {\rm Diagonal}\Big(C_\ell(\Aphiarrow=1)\,\\{-}\,C_\ell(\Aphiarrow=0)\Big).
  \label{eq:Clen}
\end{multline}
This is similar to the effect of marginalizing over an extra data component which is Gaussian and has a lensing-like power spectrum with amplitude controlled by $\DAL$, but which does not have the non-Gaussian imprint of real lensing. The similarly is only partial, however, because the correction is sometimes negative (lensing reduces power at the top of peaks in the $E$-mode power spectrum), while an extra component could only have a positive power contribution. Directly modifying the noise covariance remedies this, and can add or subtract power as long as the sum of noise and lensing-like contributions still yields a positive definite total covariance (which is the case for the range of $\DAL$ explored by the MCMC chains here).

With this modification, both non-zero $\DAL$ and non-zero $\Aphi$ can generate lensing-like power in the data. The sum of the two parameters thus gives the total lensing-like effect on the data power spectrum, and most closely matches the typical definition of the $\AL$ or $A_{\rm lens}$ parameter, which in our case is a ``derived'' parameter,
\begin{align}
  \AL = \Aphi + \DAL.
\end{align}
If no residual lensing-like power beyond the actual lensing generated by $\Aphi$ is needed to explain the data, one expects to find $\DAL\,{=}\,0$ and $\AL\,{=}\,1$. 

Because the power spectrum of the data could be just as well explained by $\DAL\,{=}\,1$ and $\Aphi\,{=}\,0$, the extent to which we infer non-zero $\Aphi$ when $\DAL$ is a free parameter confirms that not just power spectrum information is contributing to the constraint, but also quadratic $\vb{L}\,{\neq}\,0$ modes and higher-order moments. Correspondingly, marginalizing over $\DAL$ is equivalent to removing power spectrum information from the $\Aphi$ constraint, giving us the tool needed to separate out this information.

A consequence of the modification to the $\Cn$ operator in Eq.~\eqref{eq:CnAL} is that it is no longer easily factorizable in any simple basis. This presents three new numerical challenges for our MCMC chains: 1) applying the inverse of $\Cn$, 2) drawing Gaussian samples with covariance $\Cn$, and 3) computing the determinant of $\Cn$.  Inversion turns out to be fairly easily performed with a negligible $\mathcal{O}(10)$ iterations conjugate gradient. Sampling is performed by computing $\Cn^{\nicefrac{1}{2}}\,\xi$ with the same ODE-based solution used in Eq.~\eqref{eq:sqrtode}. The determinant (as a function of $\DAL$) is the most difficult piece, but can be computed utilizing the method described in \cite{fitzsimons2017}. This involves swapping the log determinant for a trace
\begin{multline}
  \log\det \left[ \Cn + \DAL \, \op{A} \, \op{C}_{\rm len} \, \op{A}^\dagger  \right] = \\
  = \sum_{k=1}^\infty \tr \left\{ \left[- \DAL \op{A} \, \op{C}_{\rm len} \, \op{A} ^\dagger \Cn \right]^k\right\} + C, \label{eq:logdetCnAL}
\end{multline}
where $C$ is a constant that is independent of $\DAL$ and can thus be ignored. The trace is then evaluated stochastically using a generalization of Hutchinson's method \citep{hutchinson1990} to complex vectors \citep{iitaka2004}, which evaluates the trace of some matrix $\op{M}$ as $\langle z^\dagger \op{M} z \rangle$ where $z$ are vectors of unit-amplitude random-phase complex numbers, here in the $EB$ Fourier domain. The summation in Eq.~\eqref{eq:logdetCnAL} converges since our matrix is positive definite, and only 20 terms are needed to give sufficient accuracy in the $\DAL$ region explored by the chain. Note also that because the powers of $\DAL$ factor out of the trace, the traces can be pre-computed once at the beginning of the chain. In terms of sampling, $\DAL$ is a ``fast'' parameter and does not significantly impact chain runtime. 

\begin{figure}
  \includegraphics[width=\columnwidth]{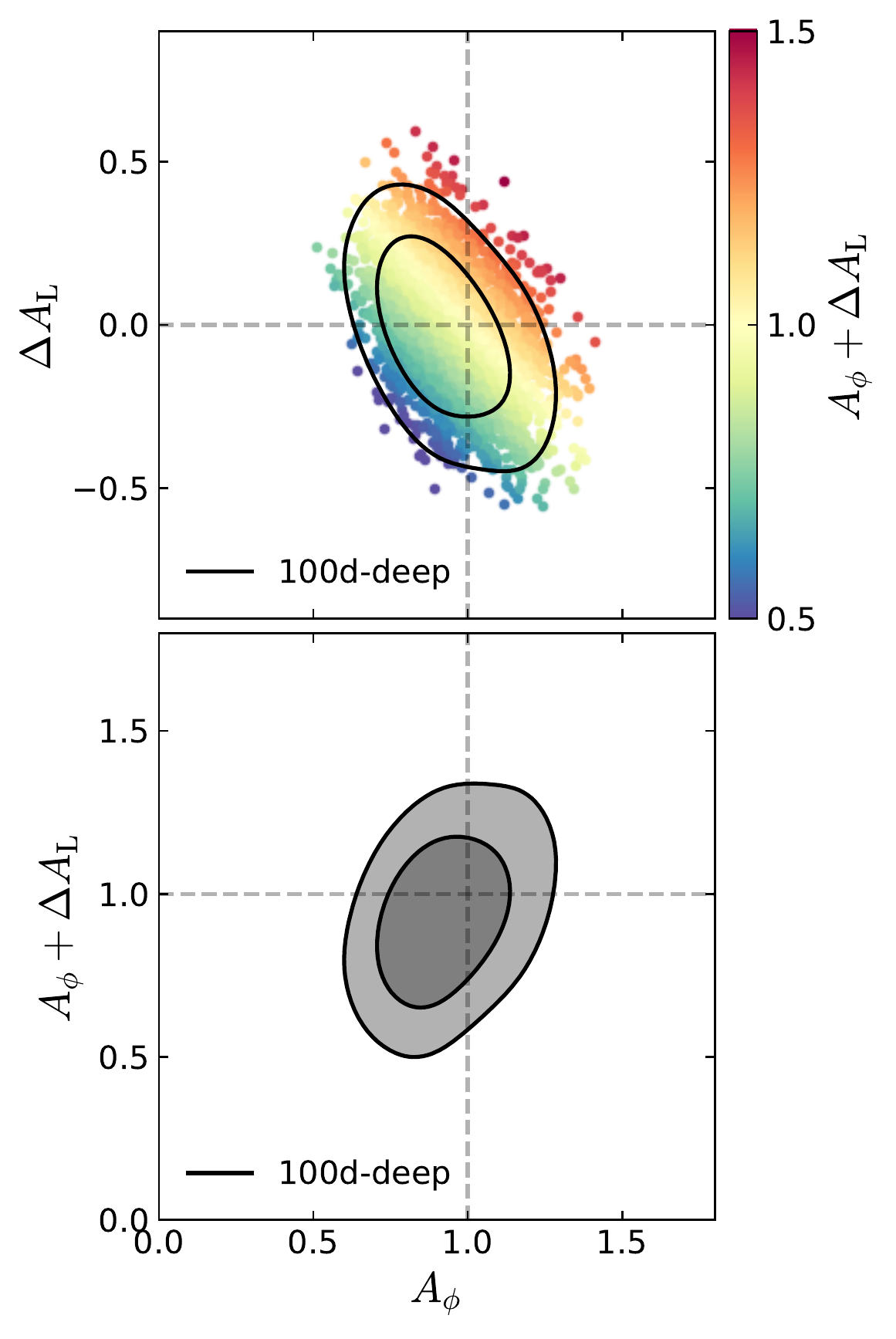}
  \caption{(Top panel) Joint constraints from the \dsultradeep dataset on the amplitude of the lensing potential, $\Aphiarrow$, and the residual lensing-like power, $\DAL$. The correlation coefficient between the two is $\rho\,{=}\,{-}0.40(5)$, demonstrating only about 9(3)\% of the $\Aphiarrow$ constraint originates from the power spectrum of the data. (Bottom panel) The same posterior as in the top panel but in terms of the $\AL=\Aphiarrow+\DAL$ parameter, which controls the total lensing-like power in the data model. These results demonstrate the unique ability of the Bayesian lensing procedure to infer parameters from an optimal lensing reconstruction and from delensed bandpowers while easily and exactly accounting for correlations between the two.}
  \label{fig:AphiAL}
\end{figure}

In the top panel of Fig.~\ref{fig:AphiAL}, we show joint constraints on $\DAL$ and $\Aphi$ from the \dsultradeep data. Here we find,
\begin{alignat}{4}
  \DAL       &=& \; &0.024(9)  &\;\pm\;& 0.170(7) \\
  \Aphiarrow &=& \; &0.955(14) &\;\pm\;& 0.135(10) \label{eq:AphiAL_Aphi}
\end{alignat}
The two parameters are visibly degenerate, with cross-correlation coefficient $\rho\,{=}\,{-}\,0.40(5)$. One can calculate by how much $\sigma(\Aphi)$ is degraded due to marginalizing over $\DAL$ as $1/\sqrt{1-\rho^2}$, which here gives a 9(3)\% degradation. Thus, relatively little information on $\Aphi$ comes from the power spectrum of the data, instead most of the constraining power originates from lensing non-Gaussianity. Because of this small impact and for simplicity, we fix $\DAL\,{=}\,0$ for the remaining results in this paper. However, we note that the 9(3)\% contribution from the power spectrum is important to keep in mind when comparing to the QE result in the next section.

The degeneracy between the two parameters arises because both $\Aphi$ and $\DAL$ modify the power spectrum of the data model in (intentionally) identical ways. The plotted samples in the top panel of Fig.~\ref{fig:AphiAL} are colored by their corresponding value of $\AL$, demonstrating that the degeneracy direction is indeed mostly aligned with $\AL$. This is consistent with the physical intuition that the total lensing-like power should be a well-constrained quantity, regardless of how much of the power is attributed to non-Gaussian lensing or not. The bottom panel shows the same posterior in terms of $\Aphi$ and $\AL$. As compared to $\Aphi$ and $\DAL$, the correlation coefficient switches sign and reduces slightly to $\rho\,{=}\,0.38(5)$. 

Correlations between $\Aphi$ and $\AL$ have been negligible in all previous lensing results from data, but are of considerable interest moving forward as it is likely they will need to be accurately quantified in the future. Previous work on this topic includes \cite{schmittfull2013a}, who computed the correlation between $\Aphi$ estimated via the QE and $\AL$ estimated via a traditional power spectrum analysis, finding at most a 10\% correlation for temperature maps at {\it Planck}-like noise levels. \cite{peloton2017a} extended similar calculations to polarization, finding correlations in the 5\%--70\% range for CMB-S4-like polarization maps, depending on the exact multipole ranges considered, if a realization-dependent noise subtraction is performed, and whether $T$, $E$, and/or $B$ are used to estimate $\AL$. The correlation is largest when using $B$, since $B$ is entirely sourced by lensing and thus contains much of the same information as $\phi$. For the \dsultradeep data, there is twice the Fisher information for $\AL$ in $B$ as compared to $E$, which means our observed correlation should be on the higher end. This is counteracted by the fact that our data is noisier than the CMB-S4 noise levels assumed in \cite{peloton2017a}, meaning we should see a lower correlation. Ultimately, although we have not repeated their calculation for our exact noise levels, our observed correlation has the same sign and reasonably agrees in amplitude with their prediction, despite the fairly different analysis.

It is useful to consider what it would take for frequentist methods such as the ones used in these previous works to reach equivalence with the Bayesian approach in terms of quantifying $\Aphi$-$\AL$ correlations, or more generally, quantifying correlations between the reconstructed lensing potential and the CMB. First, they would need to be extended beyond the QE, which would introduce computational cost and conceptually complexity. Second, they would need to be extended to compute not just correlations of the lensing reconstruction with the raw (lensed) data, but also with delensed data as well. Although not immediately obvious, this is automatically handled in the Bayesian approach. This is because, despite that the Bayesian procedure does not constrain $\AL$ by way of explicitly forming a delensed power spectrum, it exactly accounts for the actual posterior distribution of the lensed data maps. For example, if $\phi$ were perfectly known such that there were no scatter in the MCMC $\phi$ samples, this would yield no excess lensing variance when estimating $\AL$, simply an anisotropic but perfectly known lensed CMB covariance, corresponding to perfect delensing. Whether it is as easy to estimate such correlations in the frequentist approach is unclear, but we highlight the relative simplicity with which it was attained here. It required no additional costly simulations or complex analytic calculations, only the introduction of $\DAL$ into the posterior. 

Although outside of the scope of this paper, this approach can be used not just for $\DAL$ but any other cosmological parameter which controls the unlensed power spectra. It thus serves as a Bayesian analog to existing frequentist methods for parameter estimation from delensed power spectra \citep{han2020}, immediately allowing inclusion of lensing reconstruction data, and giving a path to the type of joint constraints from both that will be important for optimally inferring cosmological parameters from future data \citep{green2016}.

\subsection{Improvement over quadratic estimate}

One of the main goals of this work is to demonstrate an improvement in the Bayesian pipeline when compared to the QE result. This improvement arises because the QE ceases to be approximately minimum-variance around 5\,$\mu$K-arcmin, close to the noise levels of the \dsultradeep observations.

\begin{figure}
  \includegraphics[width=\columnwidth]{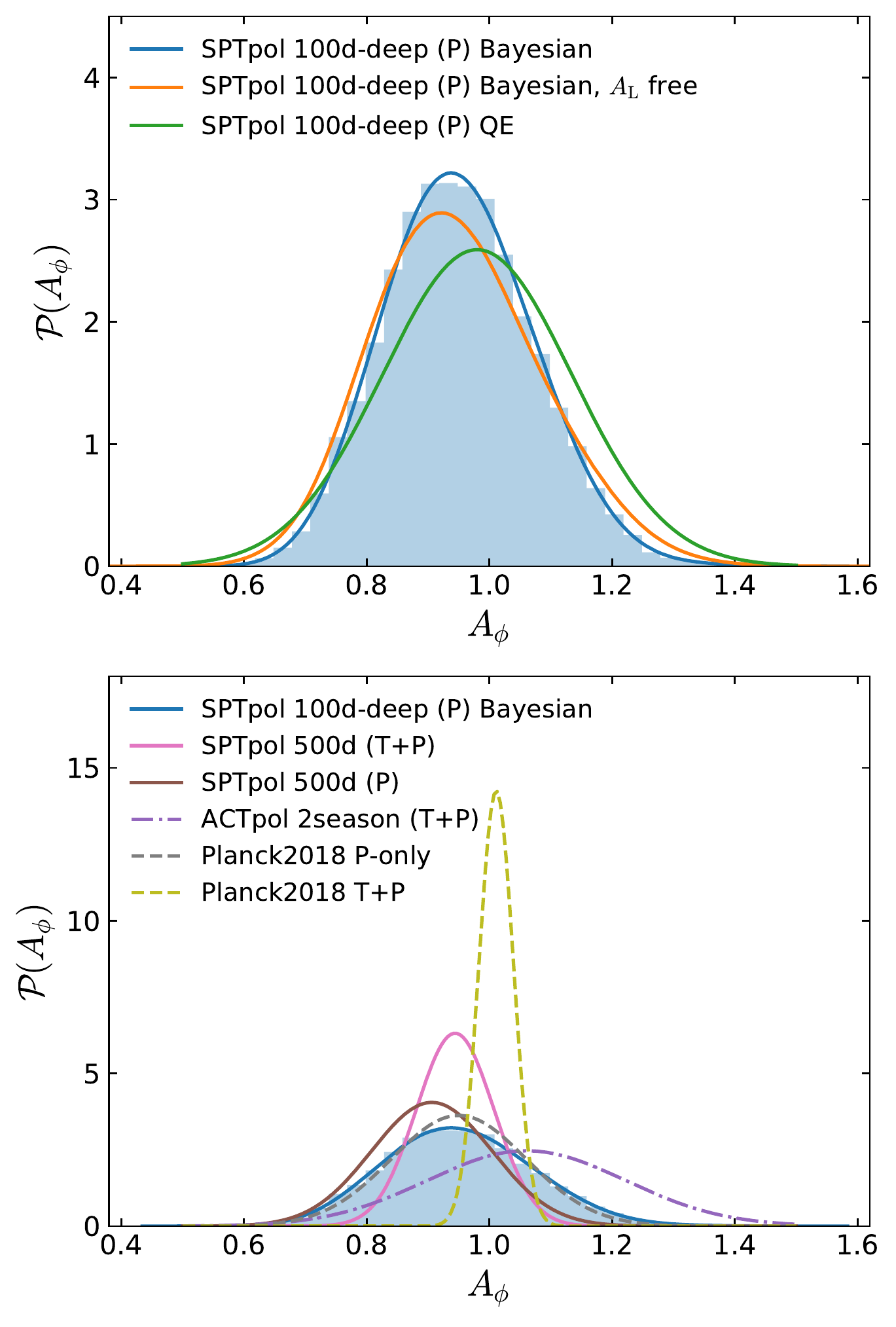}
  \caption{(Top panel) Posterior distribution of $\Aphiarrow$ as determined by the Bayesian and QE procedures. The blue bars are a histogram of the samples in the chain from the Bayesian procedure and the solid blue line is the Blackwell-Rao posterior. The orange curve removes information from the power spectrum of the data by marginalizing over $\AL$, and the green curve is the Gaussian estimate from fitting the QE bandpowers. The 17\% improvement in error bar in the $\AL$-marginalized Bayesian case over the QE is a main result of this work. (Bottom panel) Comparison of the Bayesian result with other measurements of $\Aphi$ in the literature. The result here achieves the lowest-yet effective noise level on $\phi$, although other results achieve better $\Aphi$ constraints with a larger observation region.}
  \label{fig:posterior_Aphi}
\end{figure}

The baseline \dsultradeep Bayesian constraint is
\begin{align}
  \Aphi = 0.949(8) \pm 0.122(5) \;\;\;\; {(\rm Bayesian)}
  \label{eq:ultradeep_Aphi_Bayesian}
\end{align}
For the exact same data set, the QE constraint yields
\begin{align}
  \Aphi = 0.995 \pm 0.154 \;\;\;\; {(\rm QE)}
  \label{eq:ultradeep_Aphi_QE}
\end{align}
This represents an improvement in the $1\,\sigma$ error bar of 26(5)\%, summarized in Fig.~\ref{fig:posterior_Aphi}.

The shift in the central value between the two results is $\Delta\Aphi\,{=}\,0.046(8)$. Note that these results are ``nested'' because the QE uses only quadratic combinations of the data while the Bayesian result implicitly uses all-order moments. Because of this, one can follow \cite{gratton2019} (hereafter \citetalias{gratton2019}) to calculate the standard deviation of the expected shift as $\sigma_{\Delta\Aphi}\,{=}\,(\sigma^2_{\rm QE}\,{-}\,\sigma^2_{\rm Bayesian})^{\nicefrac{1}{2}}\,{=}\,0.10(6)$. The observed shift therefore falls within the $1\,\sigma$ expectation. 

Of this improvement, we have ascertained in the previous section that 9(3)\% percent stems from the power spectrum of the data, which is not used by the QE, but could be included if we combined with traditional power spectrum constraints on $\AL$. This leaves a 17(6)\% improvement as the most fair comparison between Bayesian and QE results. To ascertain whether this is in line with expectations, we have performed a suite of generic mask-free 100\,deg$^2$ simulations with varying noise levels and $\ell_{\rm max}$ cutoffs for the reconstruction. For each of these sims, we compute the QE or joint MAP $\phi$ estimate, compute the cross-correlation coefficient, $\rho_L$, with the true $\phi$ map, then compute the effective Gaussian noise, given by $N_L^{\phi\phi} = C_L^{\phi\phi}(1/\rho_L^2-1)$. From this noise, we compute Gaussian constraints on $\Aphi$ without including the power spectrum of the data, such that these should be compared to the 17(6)\% result. Improvements in $\Aphi$ and in $N_{L=200}^{\phi\phi}$ are shown in Fig.~\ref{fig:forecast}. Near the noise levels of the \dsultradeep field, we find around a 10\% expected improvement on $\Aphi$. 

To what do we attribute the remaining ${\sim}\,7\%$ improvement we find empirically on top of this expectation? In the absence of a pixel mask, the QE is uniquely defined, e.g. as in \cite{hu02}, and it can be shown analytically that it is the minimum-variance estimator among all which are quadratic in the data. In such a mask-free case, the 10\% expected improvement of the Bayesian result arises purely through the use of higher-order moments of the data. The masked case becomes messier, because no analytic and minimum-variance result actually exists, and there is essentially no unique definition of ``the'' quadratic estimator. Instead, there are various choices and approximations which different groups are free to make at various steps in their pipeline. Although the final result can always be made unbiased at a fiducial theory model (by computing biases via Monte Carlo and subtracting them), approximations along the way introduce small amounts of extra variance. In our pipeline in particular, we take the standard approach of using the mask-free QE weights (because only the mask-free case has an analytic solution), and do not attempt to optimally spatially weight the reconstructed $\phi$ map before taking its power spectrum. Both are expected to lead to slight sub-optimality at these noise levels \citep{mirmelstein2019}, and are an expected contribution of the remaining 7\%. Finally, Monte Carlo error in the estimate of the standard deviation as well as chance realization-dependence of the posterior may contribute a few percent as well.

\subsection{Joint systematics and cosmological constraints}
\label{sec:systematics}

A unique feature of the Bayesian approach is the ability to jointly estimate cosmological and systematics parameters by simply adding free parameters to the posterior and sampling them in the chain. Here, we have added parameters for the polarization calibration, $\Pcal$, the global polarization angle calibration, $\psipol$, temperature-to-polarization monopole leakage template coefficients, $\epsQ$ and $\epsU$, and three beam eigenmode amplitudes, $\beta_1$, $\beta_2$, and $\beta_3$.

Fig.~\ref{fig:baseline_triangle} shows constraints on all of these parameters jointly with the main cosmological parameter of interest, $\Aphi$. For $\Pcal$, $\psipol$, $\epsQ$ and $\epsU$, the blue lines indicate the best-fit value obtained from the external estimation procedures described in Secs. \ref{sec:Pcal}, \ref{sec:psipol}, and \ref{sec:leak}. The chain results agrees with these in all cases, which is an important consistency check. The beam amplitude parameters, $\beta_i$, are sampled with unit Gaussian priors centered at zero. If the data is not sensitive to them, we expect the posterior is also a unit Gaussian centered at exactly zero, which is indeed what we find. 

If our main cosmological result significantly depended on knowledge of any of these systematics, we would find a correlation between these parameters and $\Aphi$. Instead, we find that no parameter is correlated at more than the 5\% level. Using the measured covariance across all parameters, $\Sigma_{ij}$, we can calculate the fractional amount by which $\sigma(\Aphi)$ decreases if the systematics were fixed to their best-fit in the \dsultradeep chain as\footnote{We could also calculate this by running a separate chain with these explicitly fixed, which we have done as a consistency check, but using $\Sigma_{ij}$ directly is easier and is less affected by Monte Carlo error.}
\begin{align}
  \sqrt{\Sigma_{11} / (\Sigma^{-1})_{11}} \lesssim 0.01,
\end{align}
where $i\,{=}\,j\,{=}\,1$ is the entry corresponding to $\Aphi$. Thus, the systematic error contribution to the Bayesian $\Aphi$ measurement is less than 1\% of the statistical error. 

Although in this paper we do not propagate any systematic errors through the QE pipeline, for some of the same data used here, this has already been done by \cite{story2015} and \cite{wu2019}. The approach there is to modify the input data, for example multiply it by $1+\sigma(\Pcal)$ to mimick a 1\,$\sigma$ error in the $\Pcal$ parameter, where $\sigma(\Pcal)$ is determined from some external calibration procedure. The resulting change to $\Aphi$ is then taken as the 1\,$\sigma$ systematic error on $\Aphi$ due to $\Pcal$, and the errors from several systematics are added in quadrature (hence assuming that they are all Gaussian and uncorrelated). For $\Pcal$, because the quadratically estimated lensing potential power spectrum depends on the fourth power of the data, the systematic error on $\Aphi$ scales as $4\,{\times}\,\sigma(\Pcal)$ to linear order, and can become significant even for modest calibration error. Indeed, using the above procedure, \cite{wu2019} found the systematic error on $\Aphi$ from polarization was nearly half of the statistical uncertainty. 

Why then is the impact of calibration errors so much smaller in the Bayesian case here? The answer is not because $\Pcal$ is more tightly constrained; $\sigma(\Pcal)$ for the \dsultradeep data is almost 1\% (see Fig.~\ref{fig:baseline_triangle}), which would translate into a 4\% error on $\Aphi$ if the above intuition held, representing about 30\% of the statistical error on $\Aphi$. Instead, the reason is that the Bayesian procedure provides an optimal inference of $\Aphi$, in particular one which automatically reduces sensitivity to $\Pcal$. As discussed in \cite{wu2019}, such a reduction is possible in the QE case as well, where one could in theory construct an alternate version of the QE with weights modified such that much of the dependence on $\Pcal$ is canceled. This would fall into a class of modifications called bias-hardened quadratic estimators, and several have been proposed to mitigate against various systematics \citep{namikawa2013,namikawa2014,sailer2020}. The advantage of the Bayesian approach is that an equivalent to bias-hardening is performed automatically (for all of the systematics parameters that we have introduced, not just $\Pcal$), and without the need to reason about the types of modifications which might be needed to mitigate each effect.

\subsection{Consistency checks}
\label{sec:consistency}

\begin{figure}
\includegraphics[width=\columnwidth]{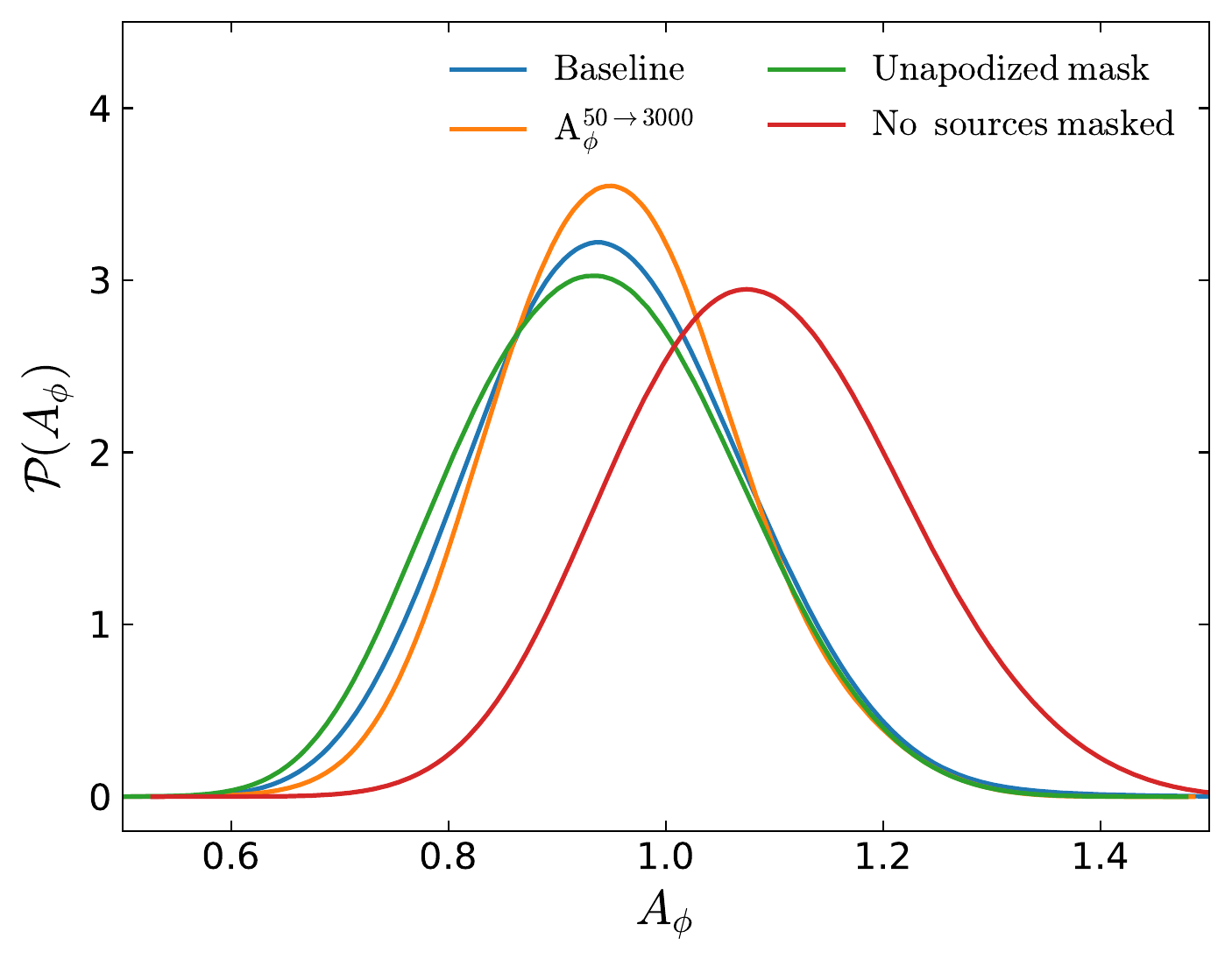}
\caption{Constraints on $\Aphi$ given various changes to the analysis as compared to the baseline result, as described in Sec.~\ref{sec:consistency}.}
\label{fig:consistency}
\end{figure}

Having presented our baseline results in the previous subsections, we now perform a number of consistency checks to see if various analysis choices have any impact on the final results. The corresponding constraints on $\Aphi$ for each case discussed here are pictured in Fig.~\ref{fig:consistency}. 

Our baseline case constrains $\Aphiarrow$. As a first check, we extend this range to encompass $\Aphi^{50\rightarrow3000}$. Here, we find
\begin{align}
  \Aphi^{50\rightarrow3000} = 0.957(8) \pm 0.114(5),
\end{align}
which is an additional 7(7)\% tighter than the baseline result, and consistent with the shift expected from \citetalias{gratton2019}. 

We next check if mask apodization has significant impact. Although the QE produces an unbiased answer regardless of mask, hard mask edges lead to larger Monte Carlo corrections and slightly larger sub-optimality of the final estimator. Conversely, the Bayesian pipeline, in theory, always produces both an unbiased and optimal result. This can be an advantage because, depending on the point source flux cut, adding a large number of apodized holes to the map can reduce the effective sky area of the observations by a non-negligible amount. One solution sometimes used in the QE case is to inpaint point source holes rather than leave them masked, and then demonstrate on simulations that negligible bias is introduced due to the inpainting \citep{benoit-levy2013,raghunathan2019}. The inpainting is often performed by sampling a constrained Gaussian realization of the CMB within the masked region, given the data just outside of the masked region. The Bayesian pipeline corresponds to simultaneously inpainting all point source holes with a different realization at each step in the MCMC chain, while accounting for the non-Gaussian statistics of the lensed CMB given the $\phi$ map at that chain step. In practice, one could imagine that the ringing created by hard mask edges induces large degeneracies in the posterior and leads to poor chain convergence. It is thus useful to verify that the Bayesian pipeline works with an unapodized mask, meaning point sources can simply be masked without apodization, and the pipeline can be used as-is without extra steps. 

To keep the apodized and unapodized cases nested, we take the original mask and set it to zero everywhere in the apodization taper. The result is the green curve in Fig.~\ref{fig:consistency}, which gives
\begin{align}
  \Aphiarrow = 0.937(15) \pm 0.124(9),
\end{align}
consistent with the \citetalias{gratton2019} expected shift. The slightly looser constraint is consistent with the unapodized case not using the data within the apodization taper, although longer chains would be needed to exactly confirm this. We do not observe a significantly worse auto-correlation length for this chain as compared to the apodized case, demonstrating that mask apodization has little effect on the Bayesian analysis.

The point source mask serves to reduce foreground contamination. Here, we have used a mask built from point sources detected in temperature, but have not attempted to cross-check if these same point sources are bright in polarization. As a simple check, we consider leaving point sources completely unmasked. In this case, we find the red curve in Fig.~\ref{fig:consistency}. This result and the baseline case are also nested. However, this time the shift in central value is inconsistent at 2.8\,$\sigma$ given \citetalias{gratton2019}. Visually inspecting the reconstructed $\kappa$ map (not pictured here) reveals obvious residuals at the locations of a few of the brightest previously masked sources. Evidently, some level of point source masking is necessary to mitigate foreground biases even in polarization. Our mask is based on a 50\,mJy flux cut in temperature. For future analyses, it will be important to determine the flux-cut which is a good trade-off between reducing foreground biases but not excising too much data.

\section{Conclusion}
\label{sec:conclusion}

We conclude with a summary of the main results along with some remarks about the Bayesian procedure and future prospects for this type of analysis. One of the main goals of this work was to apply, for the first time, a full Bayesian reconstruction to very deep CMB polarization data, and observe an improvement over the QE. This work is the second optimal lensing reconstruction ever applied to data, and the first to actually infer cosmological parameters that control the lensing potential itself. Doing so is particularly natural in the Bayesian framework, as extra parameters can always be added (sometimes trivially) and sampled over. We found a 26\% improved error bar on $\Aphi$ in the Bayesian case as compared to the QE, and a 17\% improvement after removing power spectrum information.

\begin{figure}
  \includegraphics[width=\columnwidth]{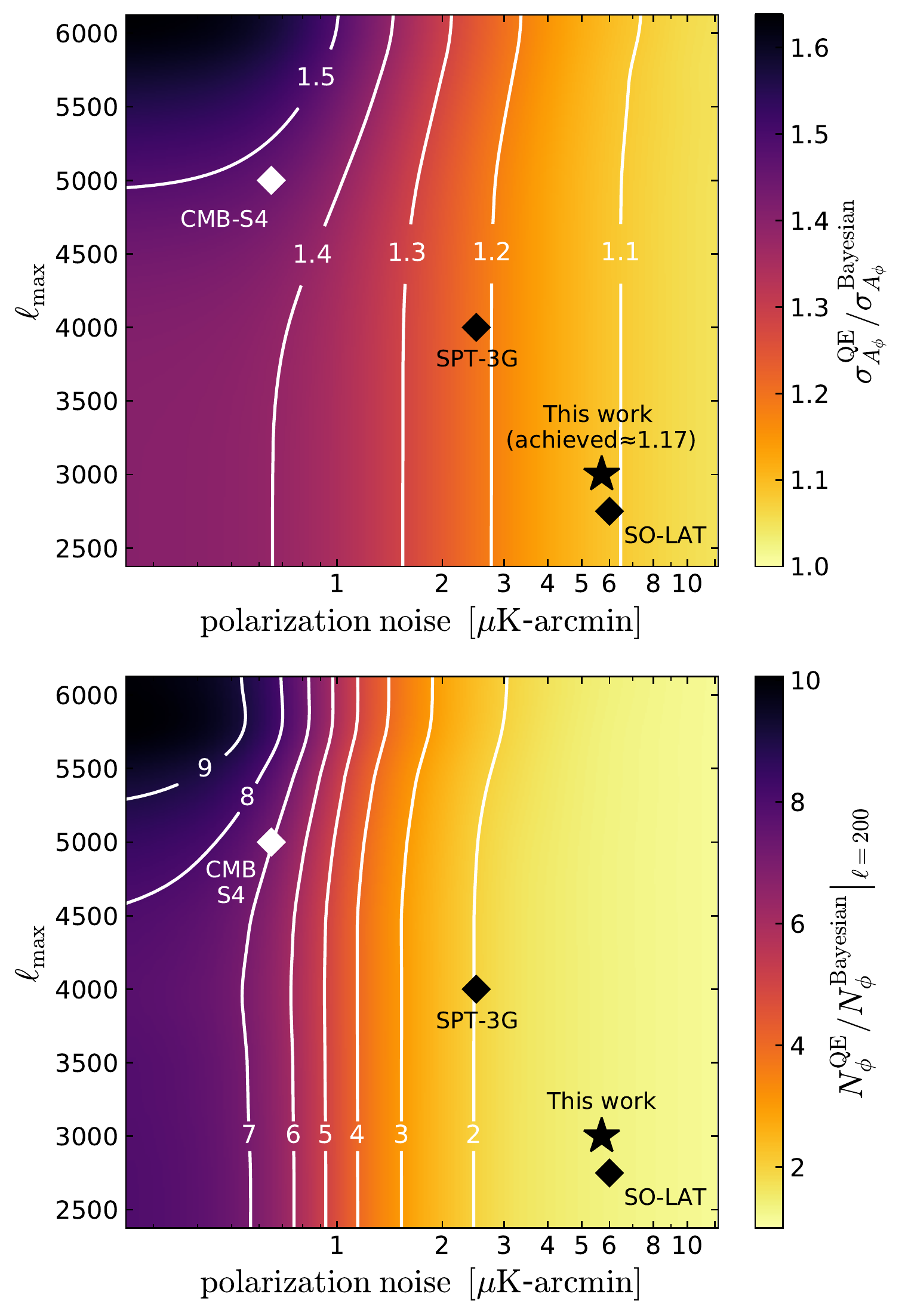}
  \caption{Forecasted improvement of Bayesian lensing reconstruction over the quadratic estimate, computed from a suite of map-level mask-free simulations. The x-axis gives the noise level in polarization and the y-axis gives the largest $\ell$ used in the reconstruction. The top panel shows the improvement in the error bar on $\Aphiarrow$. The bottom panel shows the improvement in the effective noise in the lensing reconstruction, $N_\ell^{\phi\phi}$, at $\ell\,{=}\,200$. This work achieves a slightly better improvement in $\Aphiarrow$ than predicted from these simulations due to minor sub-optimalities present in our (and typical) QE pipelines when masking and other analysis complexities exist. Forecasts for the deep CMB-S4 survey, SPT-3G, and Simons Observatory LATs are shown as diamonds. The latter lies almost directly on top of the star denoting the current work, but is offset only for visual clarity. These simulations cover roughly 100\,deg$^2$, although the relative improvements are not expected to scale appreciably with $f_{\rm sky}$.}
  \label{fig:forecast}
\end{figure}

As instrumental noise levels continue to improve in the future, we expect this relative improvement will increase. In Fig.~\ref{fig:forecast}, we forecast the relative improvement in $\Aphi$, as well more generically the relative improvement in the effective noise level of the $\phi$ reconstruction at $L\,{=}\,200$ (the choice of particular $L$ here is arbitrary, and we note that the result is only moderately sensitive to scale). By the time noise levels of the deep CMB-S4 survey are reached, the relative improvement will be around 50\% for $\Aphi$. The full story is even more optimistic, however, as $\Aphi$ is not the best parameter to reflect the lower-noise reconstruction possible in the Bayesian case. This is because once a mode becomes signal dominated, $\Aphi$ is no longer improved by further reducing the noise for that mode (only more sky can help). If we instead consider directly the effective noise level itself, which will be more indicative of the types of improvements one can achieve on parameters which are determined from noise-dominated regions of the spectra, we see that improvements of up to factors of 7 are possible. 

Another important conclusion from this work is the efficacy with which systematics can be modeled in the Bayesian framework. If a forward model for a given systematic can be devised, it can easily be included in the full posterior and estimated jointly with everything else. Doing so ensures the systematic is optimally estimated and has minimal impact on cosmological parameters. This was evident in our estimation of the $\Pcal$ parameter, which would cause a large contribution to the systematic error budget of non-bias-hardened estimators, but had negligible impact here at almost no extra work. Additionally, systematics often affect only the likelihood term, thus are ``fast'' parameters in the Gibbs sampler and more of them can be added almost for free.

Looking towards the future, the main challenges we foresee for the Bayesian approach are twofold. The first is related to a fundamental difference between Bayesian and QE (or any frequentist) method. In the frequentist case, one is free to use various approximations in the process of computing an estimator, or to null various data modes, as long as the final result is debiased (usually via Monte Carlo simulations) and this bias can be demonstrated to be sufficiently cosmology-independent. The Bayesian approach does not have any notion of debiasing, instead a forward model for the full data must be provided and guaranteed sufficiently accurate so as to ensure biases in the final answer are small. The solution we have employed here is to build the forward model with approximations to things like the transfer function, $\op{T}$, or the noise covariance, $\Cn$, which are as accurate as more sophisticated full pipeline simulations, but not prohibitive to compute at each step in the MCMC chain. Pushing to larger scales, larger sky fractions, and more complex scanning strategies will require upgrading these approximations, while keeping them fast to compute. The toolbox for these types of improvements include things like machine learning models \citep[e.g.][for a CMB application]{munchmeyer2019}, sparse operators such as the BICEP observation matrix \citep{ade2015}, or other physically motivated analytic approximations.

The second challenge of the Bayesian approach is computational. For reference, the Monte Carlo simulations needed to compute the QE here take around 10 minutes across a few hundred CPU cores. Conversely, the Bayesian MCMC chains take about 5 hours on 4 GPUs, with interpretable results returned within around an hour.\footnote{The same code can run on CPUs by switching a flag, although is factors of several slower and mainly useful for debugging.} Ignoring the mild total allocation cost of these calculations, the main difference is the longer wall-time of the MCMC chain. Since the computation is roughly dominated by FFTs, a naive scaling to e.g. the full SPT-3G 1500\,deg$^2$ footprint along with an upgraded $2^\prime$ pixel resolution (to reach scales of $\ell\,{\sim}\,5000$) gives around one week for a chain. Because the MCMC chains do not appear to require a long burn-in time, the total run-time can be reduced fairly efficiently by running more chains in parallel on more GPUs, or potentially on TPUs. Along with some planned code optimizations, we expect it will be possible to obtain results for a full SPT-3G dataset in under a day. Additionally, much of the runtime will be dominated by Wiener filtering, where our current algorithm can likely be improved, making scaling to even larger datasets possible. It may be noteworthy to highlight that the computational tools in play here, GPUs, linear algebra, and automatic differentiation, are the identical building blocks of machine learning, and are the subject of rapid technological improvements. 

The overall experience of Bayesian lensing in this work is encouraging, solving and side-stepping many difficulties which arise in other procedures. While some development is needed to extend beyond the dataset considered here, this approach appears to be a viable option for future CMB probes which will depend on methods such as these for the next generation of lensing analyses.

\begin{acknowledgements}
MM thanks Uros Seljak for useful discussions. SPT is supported by the National Science Foundation through grants PLR-1248097 and OPP-1852617.  Partial support is also provided by the NSF Physics Frontier Center grant PHY-1125897 to the Kavli Institute of Cosmological Physics at the University of Chicago, the Kavli Foundation and the Gordon and Betty Moore Foundation grant GBMF 947. This research used resources of the National Energy Research Scientific Computing Center (NERSC), a DOE Office of Science User Facility supported by the Office of Science of the U.S. Department of Energy under Contract No. DE-AC02-05CH11231. This research also used the Savio computational cluster resource provided by the Berkeley Research Computing program at the University of California, Berkeley (supported by the UC Berkeley Chancellor, Vice Chancellor for Research, and Chief Information Officer). The Melbourne group acknowledges support from the University of Melbourne and an Australian Research Council's Future Fellowship (FT150100074). Argonne National Laboratory’s work was supported by the U.S. Department of Energy, Office of High Energy Physics, under contract DE-AC02-06CH11357. We also acknowledge support from the Argonne  Center for Nanoscale Materials.

\end{acknowledgements}

\bibliographystyle{aasjournal}
\bibliography{ultradeep,marius}

\end{document}